\documentclass[useAMS,usenatbib]{mnras}

\usepackage{amsmath}	%for \text{} in math mode
\usepackage{amssymb}	%for symbols

\usepackage{graphicx}	%for including images
\usepackage{float}	%for [Hhtb] etc of figures
\usepackage{grffile}    %for using underscores in figure filenames
\usepackage{subcaption} %for having sub-captions in figures with multiple panels
\usepackage{placeins} %for \FloatBarrier, which forces all figures to be printed before continuing

\usepackage{ mathrsfs } %for slanted letters in math mode

\usepackage{hyperref} %ready-to-click referencing

\newcommand*\diff{\mathop{}\!\mathrm{d}}

\newcommand{\vect}[1]{\vec{#1}} %  ARROW vectors.

\title[Galaxy mass via lensing \& dynamics]{A spiral galaxy's mass distribution uncovered through lensing and dynamics}
\author[W.H. Trick, G. van de Ven, and A.A. Dutton]{Wilma H. Trick$^{1}$\thanks{E-mail: trick@mpia.de}, Glenn van de Ven$^{1}$, and Aaron A. Dutton$^{1,2}$\\
$^{1}$Max-Planck-Institute for Astronomy, K\"{o}nigstuhl 17, 69117 Heidelberg, Germany\\
$^{2}$New York University Abu Dhabi, PO Box 129188, Abu Dhabi, UAE\\}
\begin{document}

\date{Accepted xxx. Received xxx; in original form xxx}

\pagerange{\pageref{firstpage}--\pageref{lastpage}} \pubyear{2015}

\maketitle

\label{firstpage}

%--------------------------------------------
\begin{abstract}
We investigate the matter distribution of a spiral galaxy with a counter-rotating stellar core, SDSS J1331+3628 (J1331), independently with gravitational lensing and stellar dynamical modelling. By fitting a gravitational potential model to a quadruplet of lensing images around J1331's bulge, we tightly constrain the mass inside the Einstein radius $R_\text{ein}=(0.91\pm0.02)''$ $(\simeq1.83\pm0.04~\text{kpc})$ to within 4\%: $M_\text{ein} = (7.8\pm0.3) \times 10^{10} \text{M}_\odot$. We model observed long-slit major axis stellar kinematics in J1331's central regions by finding Multi-Gaussian Expansion (MGE) models for the stellar and dark matter distribution that solve the axisymmetric Jeans equations. The lens and dynamical model are independently derived, but in very good agreement with each other around $\sim R_\text{ein}$. We find that J1331's center requires a steep total mass-to-light ratio gradient. A dynamical model including an NFW halo (with virial velocity $v_{200} \simeq 240 \pm 40~\text{km s}^{-1}$ and concentration $c_{200} \simeq 8 \pm 2$) and moderate tangential velocity anisotropy ($\beta_z \simeq -0.4 \pm 0.1$) can reproduce the signatures of J1331's counter-rotating core and predict the stellar and gas rotation curve at larger radii. However, our models do not agree with the observed velocity dispersion at large radii. We speculate that the reason could be a non-trivial change in structure and kinematics due to a possible merger event in J1331's recent past.
\end{abstract}
%--------------------------------------------

\begin{keywords}
gravitational lensing: strong -- stars: kinematics and dynamics -- galaxies: kinematics and dynamics -- galaxies: photometry -- galaxies: structure
\end{keywords}

%--------------------------------------------
\section{Introduction} \label{sec:intro}
%--------------------------------------------

%--------------------------------------------
Determining the overall mass distribution of galaxies and separating the dark matter (DM) from the stellar mass components is a crucial step in better understanding the structure and formation of galaxies and the nature of DM.

Cosmological simulations suggest that cold DM forms cuspy halos following a Navarro-Frenk-White (NFW) profile \citep{1996ApJ...462..563N}. However, the existence of central DM density cusps in massive galaxies depends strongly on the stellar mass-to-light ratio (e.g., \citealt{2011MNRAS.416..322D}), and DM dominated dwarf galaxies even favour DM halos with cores (e.g., \citealt{1994Natur.370..629M,2001ApJ...552L..23D}). This discrepancy, known as the core-cusp problem, might be resolved by galaxy formation processes such as mergers and outflows (e.g., \citealt{2001ApJ...560..636E,2012MNRAS.421.3464P}). Especially the influence of mergers on the DM and baryonic structure of galaxies is currently an active field of research (e.g., \citealt{2009ApJ...697L..38J,2010ApJ...712...88L,2012MNRAS.425.3119H,2015MNRAS.453.2447D}).

The mass distribution of massive galaxies can be measured in completely independent fashions by gravitational lensing and dynamical modelling. Combining these two methods allows for valuable cross-checking opportunities to disentangle the galactic stellar and DM content. Massive galaxies can act as gravitational lenses, deflect the light of background sources, and give rise to multiple images. This strong gravitational lensing tightly constrains the projected total mass of the lens galaxy inside $\sim 1''$ (e.g., \citealt{2010ARA&A..48...87T}). 

The mass profile at larger galactocentric radii can be probed by gas rotation curves that directly measure the galaxy's circular velocity profile (e.g., \citealt{1980ApJ...238..471R}). However, due to its dissipative nature gas motions are very sensitive to disturbances by, e.g., spiral arms and bars (e.g., \citealt{2004dad..book.....S}). Because stars are dissipationless dynamical tracers and present almost everywhere in the galaxy, stellar dynamical modelling can complement mass constraints from lensing at small and gas motions at large radii. As stellar motions are complex---a bulk rotation around one principal axis combined with random motions in all coordinate directions---\citep{2008gady.book.....B}, full dynamical modelling of rotation, dispersion, and velocity anisotropies is needed to deduce the matter distribution.

The Sloan WFC Edge-on Late-type Lens Survey (SWELLS, WFC = Wide field camera) \citep{SWELLSI,SWELLSII,SWELLSIII,SWELLSIV,SWELLSV,SWELLSVI} is dedicated to finding and investigating spiral galaxies, which are (i) strong gravitational lenses and (ii) observed almost edge-on, such that rotation curves can be easily measured. By combining lensing and dynamical modelling, degeneracies inherent in both methods can be broken.

\begin{figure*}
\centering
\begin{subfigure}{.5\textwidth}
  \centering
  \includegraphics[width=.9\linewidth]{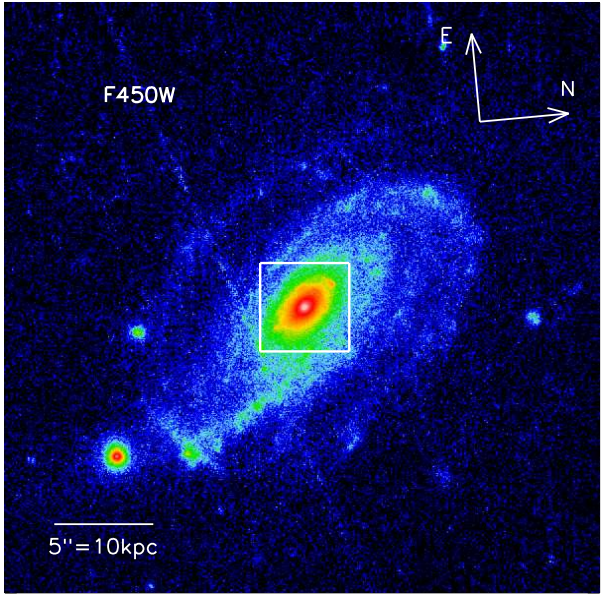}
  \caption{J1331 in the F450W filter.}
  \label{fig:F450W}
\end{subfigure}%
\begin{subfigure}{.5\textwidth}
  \centering
  \includegraphics[width=.9\linewidth]{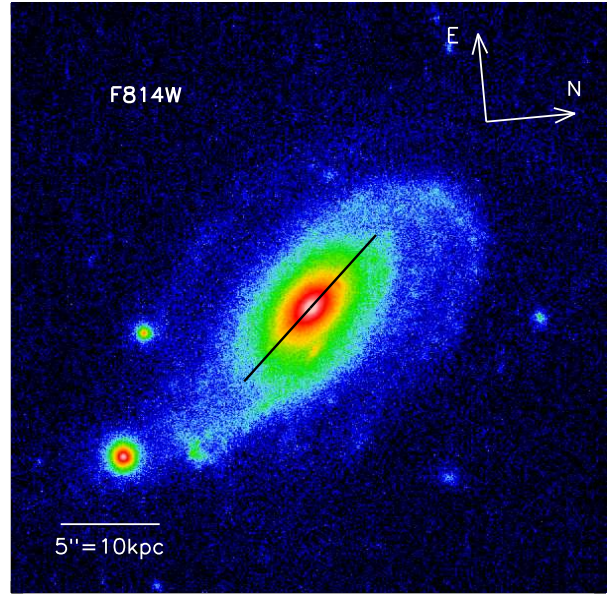}
  \caption{J1331 in the F814W filter.}
  \label{fig:F814W}
\end{subfigure}
\begin{subfigure}{.5\textwidth}
  \centering
  \includegraphics[width=.9\linewidth]{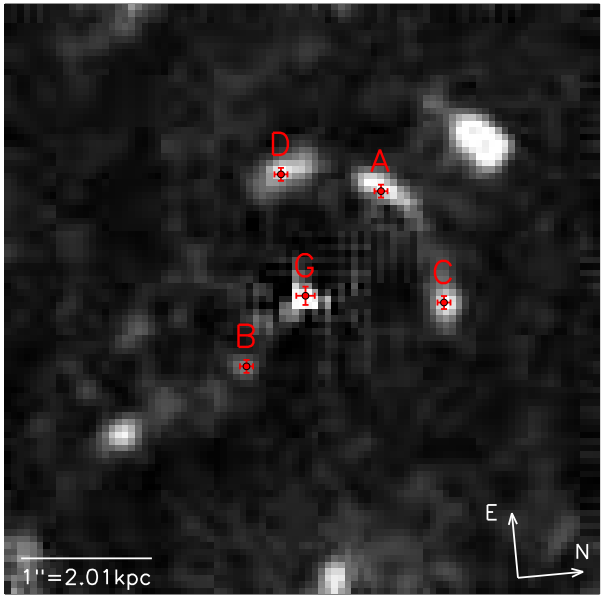}
  \caption{Lensing images around the galaxy center.}
  \label{fig:lens_just_imgpos}
\end{subfigure}%
\begin{subfigure}{.5\textwidth}
  \centering
  \includegraphics[width=.9\linewidth]{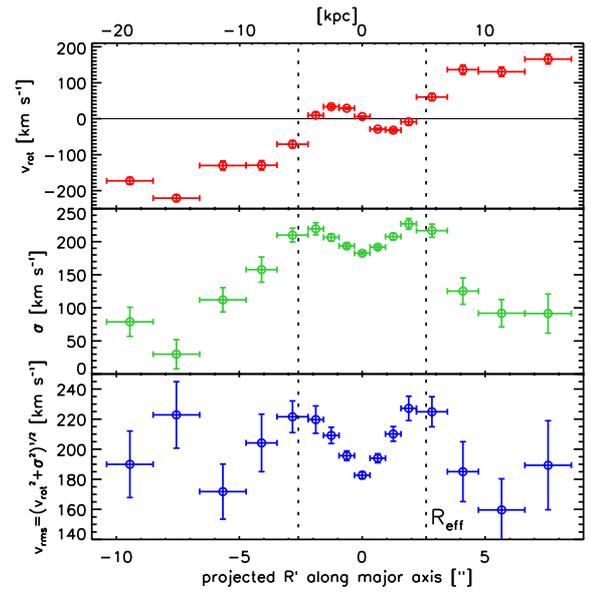}
  \caption{Stellar kinematics by \citet{SWELLSV}.}
  \label{fig:kinematics}
\end{subfigure}
\caption{Hubble Space Telescope (HST) images and stellar kinematics of the galaxy SDSS J1331+3628 (J1331), which has a large counter-rotating core and whose bulge acts as a strong lens for a bluish background source. \emph{Panel (a) and (b):} HST/WFPC2/WFC3 images of J1331 by \citet{SWELLSI} in two filters, F450W in panel (a) and F814W in panel (b). The black solid line in panel (b) shows the orientation of the major axis. Its length is $10''$ and it marks approximately the region within which we carry out the dynamical modelling. \emph{Panel (c):} Lensing images in the central region of J1331.  An IRAF \emph{ellipse} model was fitted to and then subtracted from the galaxy's F450W surface brightness in panel (a). The (smoothed) residuals within the white square in panel (a) are shown in panel (c). The four bright blobs (A, B, C, and D), which are visible in the residuals, are arranged in a typical strong lensing configuration around the center of the galaxy (G). (The configuration of the two additional blobs which lie approximately on a line with A, B, and G does not suggest that these blobs are a lensing doublet. They might rather be star-forming regions of a background galaxy.) \emph{Panel (d):} Stellar kinematics along the galaxy's major axis as measured by \citet{SWELLSV}. Shown are line-of-sight rotation velocity $v_\text{rot}$ (top), line-of-sight velocity dispersion $\sigma$ (middle) and the root mean-square (rms) velocity $v_\text{rms} = \sqrt{v_\text{rot}^2 + \sigma^2}$ (bottom). The dashed line indicates the galaxy's effective half-light radius (in the F814W filter), $R_\text{eff} = 2.6'' \hat{=} 5.2~\text{kpc}$. The $v_\text{rot}$ curve reveals that J1331 has a counter-rotating core within $R_\text{eff}$.}
\label{fig:specialJ1331}
\end{figure*}

One of the SWELLS galaxies is the massive spiral SDSS J1331+3628, to which we refer as J1331 in the remainder of this work. It has bluish spiral arms and a large reddish bulge (see Figures \ref{fig:F450W} and \ref{fig:F814W}), which is superimposed by a quadruplet of extended bluish images approximately at a distance of $1''$ from the galaxy center (see Figure \ref{fig:lens_just_imgpos}). The lensed object might be a star-forming blob of a background galaxy. J1331 stands out of the SWELLS sample because of its large counter-rotating core (see Figure \ref{fig:kinematics}), which might be an indication that J1331 underwent a merger in its recent past.

J1331 is therefore of special interest and a convenient candidate to investigate observationally if and how a merger might have modified the DM and stellar distribution of a massive spiral galaxy. This requires in particular a precise disentanglement of stellar and DM components in the galaxy's inner regions.

J1331 has already been the subject of several studies: \citet{SWELLSI} confirmed that J1331 is a strong gravitational lens, measured its apparent brightness, and estimated the stellar masses of disk and bulge. The lensing properties of J1331 were first analysed by \citet{SWELLSIII}. \citet{SWELLSV} measured the gas and stellar kinematics along the major axis, and deduced J1331's mass profile from the gas rotation curve at large radii and total mass inside the Einstein radius from gravitational lensing, focusing mostly on the outer regions of J1331.

The goal of this work is now the in-depth analysis of the matter distribution in J1331's inner regions. This will complement the previous studies of J1331 and is an important step in understanding J1331's merger-modified mass structure. We use stellar dynamical modelling in addition to lensing constraints, similar to a study of the Einstein Cross by \citet{GlennEC}. We attempt to disentangle the stellar and DM components and test if employed axisymmetric Jeans models work also in the presence of a counter-rotating core.

This paper is organized as follows: Section \ref{sec:data} summarizes the data, Section \ref{sec:Modelling} gives an overview of the modelling techniques used in this work, and Section \ref{sec:Results} presents our results on the surface photometry of J1331 using Multi-Gaussian expansions (Section \ref{sec:MGE_results}), constraints from lensing (Section \ref{sec:results_lensing}) and Jeans modelling based on the surface brightness only (Section \ref{sec:results_JAM_SB}) and including an NFW DM halo (Section \ref{sec:results_JAM_NFW}). Finally, Section \ref{sec:Discussion} uses these results to discuss J1331's possible merger history, stellar mass-to-light ratio, central kinematics, and starting points for future work.
%--------------------------------------------

%============================================
%============================================

%--------------------------------------------
\section{Data} \label{sec:data}
%--------------------------------------------

%--------------------------------------------
\subsection{Redshift and position}
%--------------------------------------------

%--------------------------------------------
J1331 is located at right ascension = 202.91800$^\circ$ and declination = 36.46999$^\circ$ (epoch J2000). \citet{SWELLSI} found from SDSS spectra that J1331 has two redshifts inside the projected radius on the sky $R'=1''$: J1331 itself has $z_{\rm d} = 0.113$, and $z_{\rm s} = 0.254$ is the redshift of the lensed background source \citep{SWELLSIII}. According to the WMAP5 cosmology \citep{WMAP5cosm}, J1331 has an angular diameter distance of 414 Mpc, which translates into a transverse scaling of $1'' \hat{=} 2.01~\text{kpc}$. We summarize all galaxy parameters also in Table \ref{tab:galaxyparameters}.
%--------------------------------------------

%--------------------------------------------
\subsection{HST imaging}
%--------------------------------------------

%--------------------------------------------
We use HST imaging of J1331 by \citet{SWELLSI}. They obtained high resolution imaging with the Hubble Space Telescope's (HST) Wide-Field Planetary Camera 2 (WFPC2) and its WF3 CCD chip. The images are a combination of four exposures with each an exposure time of $400~\text{sec}$ and were drizzled to a pixel scale of 1 pixel = $0.05''$. In particular, we use the images in the filters F450W, to identify the positions of the bluish lensing images, and F814W ($I$-band) to create a surface brightness model of the reddish bulge.
%--------------------------------------------

%--------------------------------------------
\subsection{Stellar kinematics} \label{sec:data_kinematics}
%--------------------------------------------

%--------------------------------------------
For the dynamical modelling we use the stellar kinematics for J1331 measured by \citet{SWELLSV}. They obtained long-slit spectra along J1331's major-axis with the Low Resolution Imaging Spectrograph (LRIS) on the Keck I 10m telescope. The width of the slit was $1''$ and the seeing conditions had a FWHM of $\sim 1.1''$. 1D spectra for spatial bins of different widths along the major axis were extracted. Line-of-sight stellar rotation velocities ($v_\text{rot}$) and stellar velocity dispersions ($\sigma$) were determined from the spectral region containing the absorption lines Mgb (5177 \AA) and FeII (5270, 5335 and 5406 \AA) (analogous to \citealt{SWELLSII}). Gas kinematics were extracted from Gaussian line profile fits to H$\alpha$ (6563 \AA) and [NII] (6583 and 6548 \AA) emission lines as tracers for ionized gas.

The stellar kinematics, $v_\text{rot}$, $\sigma$, and the root-mean-square velocity $v_\text{rms}^2=v_\text{rot}^2 + \sigma^2$, are shown in Figure \ref{fig:kinematics}. The rotation curve reveals a counter-rotating core within $2''\hat{=}4~\text{kpc}$. Outside of $\sim 3.5''$ there is a drop in the dispersion. This could indicate the boundary between the pressure supported bulge and the rotationally supported disk, which appears around this radius in the F450W filter in Figure \ref{fig:F450W}. However, in the brighter F814W filter in Figure \ref{fig:F814W} the smooth reddish bulge extends out to $\sim5''$. 

Inside of $\sim 4''$ the data appears to be symmetric; outside of it the data suggests that the assumption of axisymmetry seems not to be valid anymore. We subtract $2.3~\text{km s}^{-1}$ from $v_\text{rot}$ to ensure $v_\text{rot}(R'=0) \sim 0$ as a possible correction term for a misjudgement of the systemic velocity. We also symmetrize the data within $4''$ and assign a minimum error of $\delta v_\text{rms} = 5~\text{km s}^{-1}$ to the $v_\text{rms}$ data. In our dynamical Jeans modelling approach, which is based on the assumption of axisymmetry, only stellar kinematics with either $R'  \lesssim 3.5''$ or $R' \lesssim 4''$ are used. Another reason to restrict the modelling to the bulge region is that our surface brightness model (Table \ref{tab:MGEF814W} and Figure \ref{fig:MGEinnerRegions}) is only a good representation of J1331's F814W light distribution inside $\sim 5''$.
%--------------------------------------------

%--------------------------------------------
\section{Modelling} \label{sec:Modelling}
%--------------------------------------------

%--------------------------------------------
\subsection{Strong gravitational lens model} \label{sec:lensing_theo}
%--------------------------------------------

%--------------------------------------------
\subsubsection{Strong lensing formalism}
%--------------------------------------------

%--------------------------------------------
A gravitational lens is a mass distribution whose gravitational potential $\Phi$ acts as a lens for light coming from a source positioned somewhere on a plane behind the lens. We summarize the basic formalism (see also \citealt{1992grle.book.....S,1996astro.ph..6001N,2006glsw.conf....1S,2006glsw.conf...91K,2010ARA&A..48...87T}). The angular diameter distance from the observer to the lens is $D_\text{d}$, to the source plane $D_\text{s}$, and the distance between the lens and source plane is $D_\text{ds}$. The deflection potential of the lens is its potential projected along the line of sight $z'$ and rescaled to
\begin{equation}
\psi(\vect{\theta}) \equiv \frac{D_\text{ds}}{D_\text{d} D_\text{s}} \frac{2}{c^2} \int \Phi(\vect{r}=D_\text{d} \vect{\theta},z') {\ \mathrm d} z', \label{eq:psidef}
\end{equation}
where $\vect{\theta}\equiv(x',y')$ is a two-dimensional vector on the plane of the sky. The light from the source at $\vect{\beta} \equiv (x_{\rm s}',y_{\rm s}')$ is deflected according to the lens equation
\begin{equation}
\vect{\beta} = \vect{\theta}_i - \left.\vect{\nabla}_\theta \psi(\vect{\theta})\right|_{\vect{\theta}_i} \label{eq:lenseqpot}
\end{equation}
into an image $\vect{\theta}_i = (x_i',y_i')$. The gradient of the deflection potential, $\vect{\nabla}_\theta \psi(\vect{\theta})$, is equal to the angle by which the light is deflected multiplied by $D_\text{ds}/D_\text{s}$.

The inverse magnification tensor
\begin{equation}
\mathscr{M}^{-1} \equiv \frac{\partial \vect{\beta}}{\partial \vect{\theta}} \overset{(\ref{eq:lenseqpot})}{=} \left(\delta_{ij} - \frac{\partial^2 \psi}{\partial \theta_i \partial \theta_j} \right)\label{eq:magnificationtensor}
\end{equation}
describes how the source position changes with image position, the distortion of the image shape for an extended source and its magnification $\mu \equiv \text{image area}/\text{source area} = \det \mathscr{M}$. Lines in the image plane for which the magnification becomes infinite, i.e., $\det \mathscr{M}^{-1} = 0$, are called critical curves. The corresponding lines in the source plane are called caustics. The position of the source with respect to the caustic detemines the number of images and their configuration and shape with respect to each other.

The Einstein mass $M_\text{ein}$ and Einstein radius $R_\text{ein}$ are defined via the relation
\begin{equation}
M_\text{ein} \equiv M(<R_\text{ein}) \overset{!}{=} \pi \Sigma_\text{crit} R_\text{ein}^2,
\end{equation}
where 
\begin{equation}
\Sigma_\text{crit} \equiv \frac{c^2}{4\pi G} \frac{D_\text{s}}{D_\text{d} D_\text{ds}}
\end{equation}
is the critical density and $M(<R_\text{ein})$ is the mass projected along the line-of-sight within $R'=R_\text{ein}$. $M_\text{ein}$ is similar to the projected mass within the critical curve $M_\text{crit}$.
%--------------------------------------------

%--------------------------------------------
\subsubsection{Lens model} 
%--------------------------------------------

%--------------------------------------------
Following \citet{EvansWitt} we assume a scale-free model
\begin{equation}
\psi(R',\theta') = R^{'\alpha} F(\theta') \label{eq:scalefreemodel}
\end{equation}
for the lensing potential, consisting of an angular part $F(\theta')$ and a radial power-law part, with $(R',\theta')$ being again polar coordinates on the plane of the sky. $\alpha$ denotes the rotation curve slope of the lens and the case $\alpha = 1$ corresponds to a flat rotation curve. The surface density of this model is 
\begin{equation*}
\Sigma(R',\theta')= \frac{ \Sigma_\text{crit}}{2} \left(\alpha^2 F(\theta') + \frac{\partial^2}{\partial \theta^{'2}}F(\theta')\right) R^{'\alpha-2}.
\end{equation*}
We expand $F(\theta')$ into a Fourier series,
\begin{equation}
F(\theta') = \frac{a_0}{2} + \sum_{k=1}^{\infty} \left(a_k \cos(k\theta') + b_k \sin (k\theta') \right). \label{eq:Fourieransatz}
\end{equation}
For this scale-free lens model the lens equation \eqref{eq:lenseqpot} becomes
\begin{equation}
\begin{pmatrix} x_{\rm s}' \\ y_{\rm s}' \end{pmatrix} = \begin{pmatrix} R'_i \cos \theta'_i - R_i^{'\alpha-1} \left(\alpha \cos \theta'_i F(\theta'_i) - \sin \theta'_i F'(\theta'_i) \right) \\ R'_i \sin \theta'_i - R_i^{'\alpha-1} \left(\alpha \sin \theta'_i F(\theta'_i + \cos \theta'_i F'(\theta'_i) \right)\end{pmatrix}\label{eq:Fourierlenseq}
\end{equation}
\citep{EvansWitt}, where $F'(\theta') \equiv \partial F(\theta') / \partial \theta'$. When we fix the slope $\alpha$, then the lens equation is a purely linear problem and can be solved numerically for the source position $(x_{\rm s}',y_{\rm s}')$ and the Fourier parameters $(a_k,b_k)$ given one observed image at position $(x'_i=R'_i \cos \theta'_i,y'_i=R'_i \sin \theta'_i)$. 
%--------------------------------------------

%--------------------------------------------
\subsubsection{Model fitting}
%--------------------------------------------

%--------------------------------------------
The free parameters of the lens model are: the source position $(x_{\rm s}',y_{\rm s}')$, the radial slope $\alpha$ and Fourier parameters $(a_k,b_k)$ of the lens mass distribution in Equations \eqref{eq:scalefreemodel}-\eqref{eq:Fourieransatz}. We want to minimize the distance between the observed image positions, $\vect{\theta}_{{\rm o}i}$, and those predicted by the lensing model, $\vect{\theta}_{{\rm p}i}$. To avoid having to solve the lens equation \eqref{eq:Fourierlenseq} for $\vect{\theta}_{{\rm p}i}$, we follow \citet{1991ApJ...373..354K} and cast the calculation back to the source plane using the magnification tensor in Equation \eqref{eq:magnificationtensor} to approximate $\vect{\theta} \simeq (\partial \vect{\theta} / \partial \vect{\beta}) \vect{\beta} = \mathscr{M} \vect{\beta} $. The best-fitting lens model is then the one that minimizes
\begin{align}
\begin{split}
\chi^2_\text{lens} &= \sum_{i} \left|\left( \begin{matrix} \frac{1}{\Delta_x} & 0\\0 & \frac{1}{\Delta_y}\end{matrix}\right) \left( \vect{\theta}_{{\rm p}i} - \vect{\theta}_{{\rm o}i} \right)\right|^2\label{eq:chi2lens}\\
&\simeq \sum_{i} \left|\left( \begin{matrix} \frac{1}{\Delta_x} & 0\\0 & \frac{1}{\Delta_y}\end{matrix}\right)  \left.\mathscr{M}\right|_{\vect{\theta}=\vect{\theta}_{{\rm o}i}} \left( \begin{matrix} x_{\rm s}' - \tilde{x}'_{si} \\ y_{\rm s}' - \tilde{y}'_{si} \end{matrix} \right) \right|^2,
\end{split}
\end{align}
where $(\Delta_x,\Delta_y)$ are the measurement errors of the image positions $\vect{\theta}_{{\rm o}i}$. $\left.\mathscr{M}\right|_{\vect{\theta}=\vect{\theta}_{{\rm o}i}}$ is the magnification tensor and $(\tilde{x}'_{si},\tilde{y}'_{si})$ the source position according to the lens equation, both evaluated at the position of the $i$-th lensing image, $\vect{\theta}_{{\rm o}i}$. Following \citet{GlennEC} we add a term
\begin{equation}
\chi^2_\text{shape} = \lambda \sum_{k \geq 3} \frac{\left(a_k^2 +b_k^2 \right)}{a_0^2} \label{eq:chi2shape}
\end{equation}
(with $\lambda$ being some weight factor) which forces the shape of the mass distribution to be close to an ellipse. The total $\chi^2$ to minimize is therefore
\begin{equation}
\chi^2 = \chi^2_\text{lens} + \chi^2_\text{shape} \label{eq:chi2total}
\end{equation}
We set $a_1 = b_1 = 0$, which corresponds to the choice of origin; in this case the center of the galaxy.

To be able to constrain the slope $\alpha$ we would need flux ratios for the images as in \citet{GlennEC}. But the extent of the images, possible dust obscuration and surface brightness fluctuations due to microlensing events, as well as the uncertainty in surface brightness subtraction, make flux determination too unreliable and we do not include them in the fitting.
%--------------------------------------------

%--------------------------------------------
\subsection{Surface brightness model} \label{sec:MGE_theo}
%--------------------------------------------

%--------------------------------------------
\subsubsection{Multi-Gaussian Expansion (MGE)}
%--------------------------------------------

%--------------------------------------------
MGEs are used to parametrize the observed surface brightness (or projected total mass) of a galaxy as a sum of $N$ two-dimensional, elliptical Gaussians \citep{1991ApJ...366..599B,1992A&A...253..366M,1994A&A...285..723E,1999MNRAS.303..495E}. This work makes use of the algorithm and code\footnote{The IDL code package for fitting MGEs to images by \citet{Cap02} is available online at \url{http://www-astro.physics.ox.ac.uk/~mxc/software}. The version from June 2012 was used in this work.} by \citet{Cap02}. We assume all Gaussians to have the same center and position angle $\phi$, i.e., the orientation of the Gaussians' major axis measured from North through East in the polar coordinate system $(R',\theta')$ on the plane of the sky. Then a surface brightness model can be expressed as
\begin{align}
I(R',\theta') &= \sum_{i=1}^{N} I_{0,i} \exp\left[ - \frac{1}{2\sigma_i^2} \left(x^{'2} + \frac{y^{'2}}{q_i^{'2}}\right)\right]\label{eq:MGEgeneral}\\
\text{with } I_{0,i} &= \frac{L_i}{2\pi \sigma_i^2 q'_i}\label{eq:centralItotalL}\\
\begin{split}
\text{and } x'_i &= R' \cos(\theta' - \phi)\\
y'_i &= R' \sin(\theta' - \phi),
\end{split}
\end{align}
where $I_{0,i}$ is the central surface brightness of each Gaussian, $L_i$ its total luminosity, $\sigma_i$ its dispersion along the major axis and $q'_i$ the axis ratio between the elliptical Gaussian's major and minor axis.
%--------------------------------------------

%--------------------------------------------
\subsubsection{Convolution with the point spread function (PSF)}
%--------------------------------------------

%--------------------------------------------
We can also expand the telescope's PSF as a sum of circular Gaussians,
\begin{equation}
\text{PSF}(x',y') = \sum_j \frac{G_j}{2 \pi \delta_j^2} \exp\left[- \frac{1}{2 \delta_j^2} \left({x'}^2 + {y'}^2 \right)\right], \label{eq:PSFgeneral}
\end{equation}
where $\sum_j G_j = 1$, and $\delta_j$ are the dispersions of the circular PSF Gaussians. 

The observed surface brightness distribution is a convolution of the intrinsic surface brightness in Equation \eqref{eq:MGEgeneral} with the PSF in Equation \eqref{eq:PSFgeneral}: $(I \ast \text{PSF}) (x',y')$ is then again a sum of Gaussians and can be directly fitted to an image of the galaxy in question.
%--------------------------------------------

%--------------------------------------------
\subsubsection{Deprojection} \label{sec:MGE_theo_deprojection}
%--------------------------------------------

%--------------------------------------------
$I(R',\theta')$ describes the intrinsic and 2D projected light distribution or surface density of the galaxy. Under the assumption that the galaxy is oblate and axisymmetric, and given the inclination angle $\iota$ of the galaxy with respect to the observer, MGEs allow an analytic deprojection of the two-dimensional MGE to get a three-dimensional axisymmetric light distribution or density $\nu(R,z)$ for the galaxy,
\begin{equation}
\nu(R,z) = \sum_i \nu_{0,i} \exp \left[-\frac{1}{2\sigma_i}\left(R^2 + \frac{z^2}{q_i^2} \right) \right], \label{eq:deprojMGE}
\end{equation}
where $R$ is the distance from the galaxy's short axis and $z$ the height above the galactic plane.\footnote{$(R,z)$ are cylindrical coordinates aligned with the galactic plane. The primed $(R',z')$ denote polar coordinates on the plane of the sky.} The flattening $q_i$ of each axisymmetric 3D Gaussian and its central density $\nu_{0,i}$ follow from the observed 2D axis ratio $q'_i$ and surface density $I_{0,i}$ as
\begin{eqnarray}
q_i^2 &=& \frac{q_i'^2 - \cos^2 \iota}{\sin^2 \iota}\\
\nu_{0,i} &=& \frac{q_i' I_{0,i}}{q_i \sqrt{2 \pi \sigma_i^2}}.
\end{eqnarray}
%--------------------------------------------

%--------------------------------------------
\subsection{Axisymmetric dynamical model} \label{sec:model_JAM}
%--------------------------------------------

%--------------------------------------------
\subsubsection{Jeans Anisotropic Models (JAM)} \label{sec:model_JAM_JAM}
%--------------------------------------------

%--------------------------------------------
JAM modelling assumes galaxies to be (i) collisionless, (ii) in a steady state, and (iii) axisymmetric. With these assumptions the axisymmetric Jeans equations follow from the vector-valued first moment of the collisionless Boltzmann equation (see Equations (4.221)-(4.222) in \citet{2008gady.book.....B}). The Jeans equations are functions of the second velocity moments $\langle v_i v_j\rangle$ (with $i,j\in(R,z,\phi)$), the number density of (stellar) tracers $n(\vect{x})$, and the galaxy's gravitational potential $\Phi(\vect{x})$ generated by the mass density $\rho(\vect{x})$.

To be able to solve the Jeans equations, additional assumptions about the velocity ellipsoid tensor $\langle v_i v_j\rangle$ have to be made. We follow \citet{Cap08} and assume first that the galaxy's velocity ellipsoid is aligned with the cylindrical coordinate system, i.e., $\langle v_R v_z\rangle = \langle v_z v_\phi \rangle = \langle v_\phi v_R \rangle = 0$. Secondly, we assume a constant ratio between the radial and vertical second velocity moments, 
\begin{equation}
\beta_z \equiv 1 - \langle v_z^2 \rangle / \langle v_R^2\rangle, \label{eq:bz}
\end{equation}
where $\beta_z$ is the velocity anisotropy parameter.

The JAM modelling approach by \citet{Cap08} expresses the tracer and mass density in terms of MGEs (see also \citealt{1994A&A...285..723E}). The tracer density $n(\vect{x})$ is assumed to be proportional to the observed and deprojected brightness distribution $\nu(R,z)$ in Equation \eqref{eq:deprojMGE}. The mass density $\rho(R,z)$ can consist of several sets of MGEs, describing stellar and DM components. The MGE for $\Phi(R,z)$ is generated from the mass density MGE by integrating the Poisson equation \citep{1994A&A...285..723E}. 

In this way one can now calculate an unambiguous model prediction for the velocity dispersion tensor $\langle v_i v_j \rangle(R,z) = \langle v_i^2 \rangle(R,z)$ (with $i,j \in \{ R,\phi, z\}$ and $\langle v_i v_j \rangle=0$ if $i\neq j$). To compare it with observations, $\langle v_i^2 \rangle(R,z)$ has to be rotated by the inclination angle $\iota$ to the coordinate system of the observer; $(x',y')$ is the plane of the sky and $z'$ the line-of-sight, where $x'$ is aligned with the galaxy's major axis. Taking a light-weighted projection along the line-of-sight gives a model prediction for the line-of-sight velocity second moment $\langle v_\text{los}^2\rangle(x',y')$, which is comparable to actual spectroscopic measurements of the second velocity moment. Details of the derivation of $\langle v_\text{los}^2\rangle(x',y')$ from the Jeans equations using the MGE formalism are laid out in \citet{Cap08} and in the appendix of \citet{GlennEC}. The result for $\langle v_\text{los}^2\rangle(x',y')$ in particular is given in Equation (28) and Equation (A18), respectively.

The JAM modelling code\footnote{The IDL code package for Jeans Anisotropic Models (JAM) by \citet{Cap08} is available online at \url{http://www-astro.physics.ox.ac.uk/~mxc/software}. The version from June 2012 was used in this work.} by \citet{Cap08} evaluates this expression for $\langle v_\text{los}^2\rangle(x',y')$ numerically for a given luminous tracer and mass distribution MGE and a given inclination.
%--------------------------------------------

%--------------------------------------------
\subsubsection{Data comparison} \label{sec:model_JAM_compare}
%--------------------------------------------

%--------------------------------------------
As data we use stellar line-of-sight rotation velocities $v_\text{rot} = \langle v_\text{los} \rangle_\text{obs}$ and velocity dispersions $\sigma$ as described in Section \ref{sec:data_kinematics}. The JAM models give a prediction for the second line-of-sight velocity moment $\langle v_\text{los}^2 \rangle_\text{obs}$. The root-mean-square (rms) line-of-sight velocity $v_\text{rms}$ allows a data-model comparison by relating theses velocities according to 
\begin{equation}
 v_\text{rms}^2 = v_\text{rot}^2 + \sigma^2  \stackrel{!}{=} \langle v_\text{los}^2 \rangle_\text{obs}.
\end{equation}
The model derived from the Jeans equations as outlined in Section \ref{sec:model_JAM_JAM} predicts the intrinsic $\langle v_\text{los}^2 \rangle_\text{intr}\equiv\langle v_\text{los}^2\rangle$ at a given position on the sky, which needs then to be modified to $\langle v_\text{los}^2 \rangle_\text{obs}$ according to the mode of observation to be comparable to the measurements. The measured $v_\text{rms}$ in this work is a light-weighted mean for a pixel along the long-slit of the spectrograph, with width $L_y = 1''$ \citep{SWELLSV} and a certain given extent along the galaxy's major axis, $L_x$, i.e., for a rectangular aperture
\begin{equation}
\text{AP}(x',y') \equiv \left\{ \begin{array}{ll} 1 & \text{for } -\frac{L_x}{2} \leq x' < + \frac{L_x}{2}\\
& \text{and } - \frac{L_y}{2} \leq y' \leq + \frac{L_y}{2}  \\ 0 & \text{otherwise.} \end{array} \right.
\end{equation}
The light arriving at the spectrograph itself was subject to seeing, i.e., a Gaussian with Full Width Half Maximum (FWHM) of $1.1''$ \citep{SWELLSV},
\begin{equation}
\text{PSF}(x',y')\equiv\mathscr{N}(0,\sigma_\text{seeing}=\text{FWHM}/2\sqrt{2\ln2}).
\end{equation}
The model predictions for $\langle v_\text{los}^2 \rangle_\text{intr}$ have therefore to be convolved with the convolution kernel
\begin{eqnarray}
K(x',y') &\equiv& (\text{PSF} \ast \text{AP})(x',y') \nonumber\\
&=& \frac{1}{4} \prod_{u \in \{x',y'\}} \left[\text{erf}\left( \frac{L_u/2 - u}{\sqrt{2}\sigma_\text{seeing}}\right) \right.\nonumber\\
 && + \left. \text{erf} \left( \frac{L_u/2 + u}{\sqrt{2} \sigma_\text{seeing}} \right) \right],
 \end{eqnarray}
and weighted by the surface brightness $I(x',y')$, i.e.,
\begin{eqnarray}
I_\text{obs} &=& I \ast K,\\
\langle v_\text{los}^2 \rangle_\text{obs} &=& \frac{(I \langle v_\text{los}^2\rangle_\text{intr}) \ast K}{I_\text{obs}}.
\end{eqnarray}
We modified the JAM code by \citet{Cap08} to use this convolution kernel with non-square pixels. The JAM code then performs the convolution numerically. We set $L_x = 0.21''$ as the width of the model pixel, and get a prediction for the actual measurements in bins of width $0.63''$, $1.26''$ and $1.89''$ \citep{SWELLSV} as light-weighted mean from each 3, 6 and 9 model pixels.
%--------------------------------------------

%--------------------------------------------
\subsubsection{Rotation curve} \label{sec:model_JAM_rotation}
%-------------------------------------------- 

%--------------------------------------------
The intrinsic galaxy rotation curve is the first velocity moment $\langle v_\phi\rangle = \sqrt{\langle v_\phi^2 \rangle - \sigma_\phi^2}$. The observed rotation velocity is the projection of the light-weighted contributions to $\langle v_\phi\rangle$ along the line-of-sight (Equation (31) in \citet{Cap08}).

The first velocity moments cannot be uniquely determined from the Jeans equations, which give only a prediction for the second velocity moments. Further assumptions are needed to separate the second velocity moments into ordered and random motion. \citet{Cap08} assumes that in a steady state there is no streaming velocity in $R$ direction, i.e., $\langle v_R \rangle = 0$, and therefore $\sigma_R^2 = \langle v_R^2 \rangle$. Then \citet{Cap08} relates the dispersions in $R$ and $\phi$ direction such that
\begin{equation}
\langle v_\phi\rangle = \sqrt{\langle v_\phi^2 \rangle - \sigma_\phi^2} \equiv \kappa \sqrt{\langle v_\phi^2 \rangle - \langle v_R^2 \rangle},
\end{equation}
and the $\kappa$ parameter quantifies the rotation: $\kappa = 0$ means no rotation at all, and $|\kappa| = 1$ describes a velocity dispersion ellipsoid that is a circle in the $R$-$\phi$ plane \citep{Cap08}. The sign of $\kappa$ determines the rotation direction. We can assign a constant $\kappa_i$ to every Gaussian in the MGE formalism and calculate the light-weighted circular velocity curve, given the second velocity moments found from the Jeans equations (see Equations (37) and (38) in \citet{Cap08} for the intrinsic and observed rotation curves, respectively).

To model the counter-rotating core of J1331 with one free parameter, we employ the condition that the overall $\kappa(R)$ profile should smoothly and relatively steeply transit from $\kappa(R) = -\kappa' < 0$ at small $R$ through $\kappa(R_0) = 0$ and increase to $\kappa(R) = \kappa' > 0$ at large $R$. Our imposed profile is
\begin{equation}
\kappa(R) \equiv \kappa' \frac{R^2 - R_0^2}{R^2 + R_0^2}. \label{eq:kappa_profile}
\end{equation}
We find $\kappa'$ by matching the model $\langle v_\text{los} \rangle_\text{obs}$ with the symmetrized $v_\text{rot}$ data, where for a given $\kappa'$ the $\kappa_i$ are found from fitting the MGE generated profile $\kappa(R) = \sum_i \kappa_i \nu_i(r)/\sum_i \nu_i(r)$ to Equation \eqref{eq:kappa_profile}. The observed zero-point is at $R'_0\approx 2''$. In the deprojected galactic plane the radius of zero rotation would be at a $R_0 \gtrsim 2''$, and we choose it to be at $2.2''$.
%--------------------------------------------

%--------------------------------------------
\subsubsection{Including an NFW halo} \label{sec:model_JAM_NFW}
%--------------------------------------------

%--------------------------------------------
As mentioned above, JAM modelling allows us to incorporate an invisible matter component in addition to the luminous matter in the form of an MGE. In Section \ref{sec:results_JAM_NFW}, we will include a spherical Navarro-Frenk-White (NFW) DM halo \citep{1996ApJ...462..563N} in the dynamical model. The classical NFW profile has the form
\begin{equation}
\rho_\text{NFW}(r) \propto r^{-1} \left( r+r_{\rm s} \right)^{-2} \label{eq:NFWprofile}
\end{equation}
and two free parameters, the scale length $r_{\rm s}$ and a parameter describing the total mass of the halo. We will use $v_\text{200}\equiv \sqrt{GM_{200}/r_{200}}$, which is the circular velocity at the radius $r_\text{200}$ within which the mean density of the halo is 200 times the cosmological critical density $\rho_\text{crit} \equiv (2H^2)/(8\pi G)$, and where $M_{200} \equiv M(<r_{200}) \stackrel{!}{=} \frac 43 \pi r_{200}^3 \times 200\rho_\text{crit}(z=0)$, with $\rho_\text{crit}(z=0)=1.43 \times 10^{-7} \text{M}_\odot / \text{pc}^3$ in the WMAP5 cosmology by \citet{WMAP5cosm}.
The mass concentration of the NFW halo is quantified by $c_{200}\equiv r_{200} / r_{\rm s}$.
There is a close relation between the concentration and halo mass in simulations \citep{1996ApJ...462..563N}. \citet{Maccio08} found this relation for the WMAP5 cosmology to be
\begin{equation}
\langle \log c_{200} \rangle (M_{200}) = 0.830 - 0.098 \log \left(h \frac{M_{200}}{10^{12} \text{M}_\odot} \right) \label{eq:Maccio08}
\end{equation}
(their Equation 10), with a Gaussian scatter of $\sigma_{\log c_{200}} = 0.105$ (their Table A2). In Section \ref{sec:results_JAM_NFW}, we will use this relation as prior information to guide the modelling towards realistic NFW halo shapes.
%--------------------------------------------

%--------------------------------------------
\section{Results} \label{sec:Results}
%--------------------------------------------

%============================================
\begin{table}
\centering
\caption{General galaxy parameters of J1331.}
\begin{tabular}{lllr}
\hline
redshift \citep{SWELLSIII}                 & $z_{\rm d}$ & $0.113$ \\
angular diameter distance & $D_\text{d}$ [Mpc] & $414$ \\
scaling                   & $1~\text{kpc} / 1''$ & $2.006$ \\
position angle from North          & $\phi$ [$^\circ$] & $42.90$\\
average axis ratio & $q'$ & $0.598$\\
average ellipticity & $\epsilon = 1 - q'$ & $0.402$ \\
estimated inclination & $\iota$ [$^\circ$] & $70$\\
apparent $I$-band magnitude & $m_\text{I}$ [mag] & $15.77$ \\
total $I$-band luminosity & $L_\text{I,tot}$ [$10^{10} L_\odot$] & $5.6$ \\
effective half-light radius & $R_\text{eff}$ [$''$] & $2.6$ \\
& $R_\text{eff}$ [pc]& 5.2 & \\
\hline
\end{tabular}
\label{tab:galaxyparameters}
\end{table}
%============================================

%--------------------------------------------
\subsection{Surface photometry for J1331 with MGEs} \label{sec:MGE_results}
%--------------------------------------------

%--------------------------------------------
In this section we construct a model for J1331's intrinsic light distribution in terms of MGEs (see Section \ref{sec:MGE_theo}). We use the HST image in the F814W ($I$-band) filter (Figure \ref{fig:F814W}) because J1331's central stellar component appears at longer wavelengths (i) smoother and more extended than in the F450W filter (Figure \ref{fig:F450W}), as it is less sensitive to young clumpy star-forming regions, (ii) much brighter than the bluish lensing images, and (iii) the imaging is less prone to extinction.

%============================================
\begin{table}
\centering
\caption{F814W PSF MGE: Parameters of the circular 4-Gaussian MGE in Equation \eqref{eq:PSFgeneral} fitted to the radial profile of the synthetic HST/F814W PSF image.}
\begin{tabular}{ccc}
\hline
$j$ & $G_j$ & $\delta_j$ [$''$] \\\hline
1 & 0.184 & 0.038\\
2 & 0.485 & 0.085\\
3 & 0.222 & 0.169\\
4 & 0.109 & 0.487\\\hline
\end{tabular}
\label{tab:PSFMGEF814W}
\end{table}
%============================================

%--------------------------------------------
\subsubsection{PSF for the HST/F814W filter}
%--------------------------------------------

%--------------------------------------------
The one-dimensional MGE in Equation \eqref{eq:PSFgeneral} is fitted to the radial profile of a synthetic image of the HST/F814W filter PSF, ignoring diffraction spikes and using the code by \citet{Cap02}. The MGE parameters of the normalized PSF model are given in Table \ref{tab:PSFMGEF814W}.
%--------------------------------------------

%============================================
\begin{figure*}
\centering
\begin{subfigure}{.5\textwidth}
  \centering
  \includegraphics[width=.8\columnwidth]{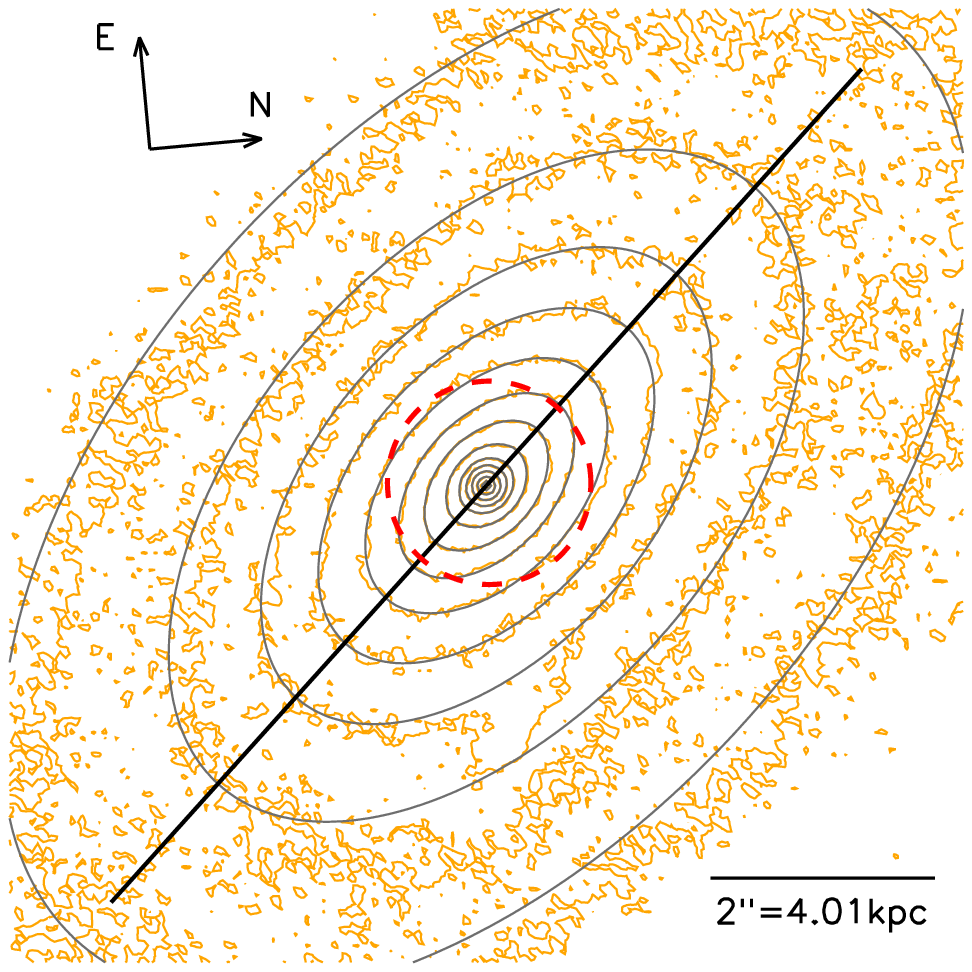}
  \caption{MGE for J1331's inner regions.}
 \label{fig:MGEinnerRegions}
\end{subfigure}%
\begin{subfigure}{.5\textwidth}
  \centering
  \includegraphics[width=.8\columnwidth]{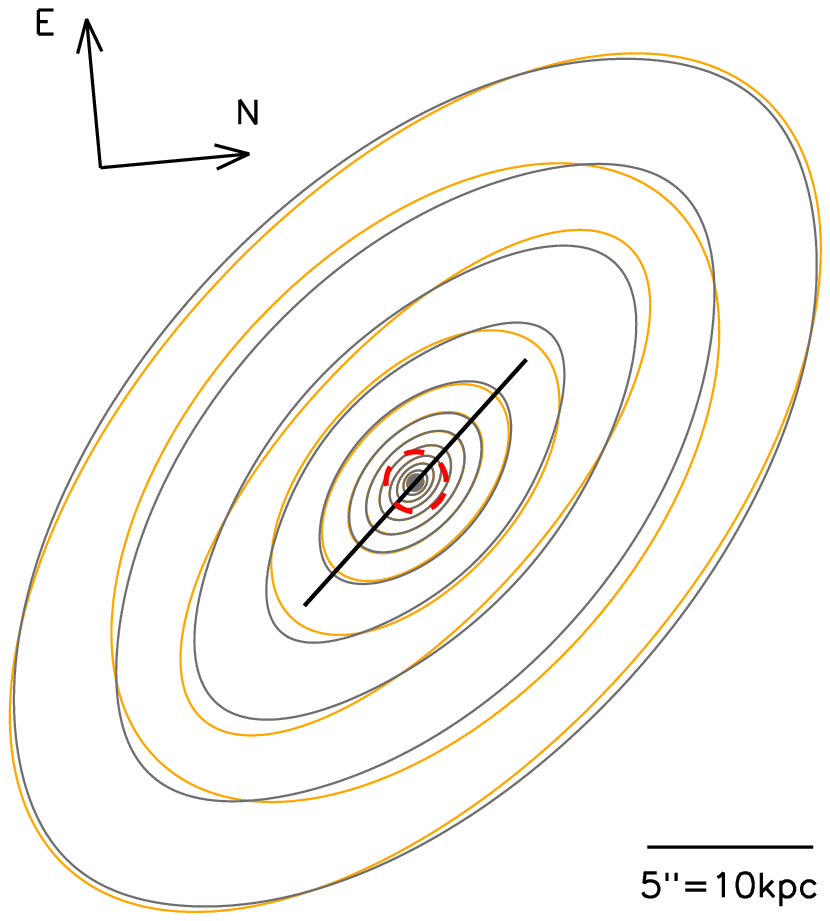}
  \caption{MGE and IRAF \emph{ellipse} model for J1331's outer regions.}
 \label{fig:MGEouterRegions}
\end{subfigure}%
\caption{MGEs for J1331's surface brightness distribution. Comparison of contours with constant F814W surface brightness (orange lines) with the corresponding iso-brightness contours of the best-fitting MGE, convolved with the PSF in Table \ref{tab:MGEF814W}, (grey lines). The black line has a length of $10''$ and its orientation corresponds to the galaxy's position angle as found in Table \ref{tab:galaxyparameters}. For comparison, the Einstein radius as found in Table \ref{tab:bestfitlensmodel} is indicated as red dashed circle. \emph{Panel (a):} Central regions of J1331. The MGE model is a good representation of the galaxy's light distribution along the major axis within $\sim 5''$. Its parameters are given in Table \ref{tab:MGEF814W}. This MGE is used as model for the stellar tracer distribution in the dynamical Jeans modelling in Sections \ref{sec:results_JAM_SB} and \ref{sec:results_JAM_NFW}. \emph{Panel (b):} Outer regions of J1331. The orange lines indicate here contours of a smooth IRAF \emph{ellipse} fit to J1331 in the F814W filter; the grey lines are the corresponding best-fitting MGE. This MGE is not used for dynamical model fitting because the dynamics in the outer regions are strongly affected by non-axisymmetries (e.g., spiral arms). We use it, however, to estimate the galaxy's total luminosity and effective radius, and to get an estimate for the dynamics of the outer regions.}
\end{figure*}
%============================================

%============================================
\begin{table*}
\centering
\caption{Parameters of the best-fitting MGE to the F814W surface brightness of J1331 in Figure \ref{fig:MGEinnerRegions}. The fit is best inside a radius of $5''$. The position angle is given in Table \ref{tab:galaxyparameters}. This MGE is used in the dynamical modelling in Sections \ref{sec:results_JAM_SB} and \ref{sec:results_JAM_NFW}. The first column gives for each Gaussian the total F814W luminosity in Equation \eqref{eq:centralItotalL} in units of counts. The second column is the corresponding $I$-band peak surface brightness in Equation \eqref{eq:MGEgeneral} in units of a luminosity surface density. The third and fourth column give the dispersion and the last column the axis ratio of the Gaussian in Equation \eqref{eq:MGEgeneral}.}
\begin{tabular}{cccccc}
\hline
 & total luminosity  & surface density & \multicolumn{2}{c}{Gaussian dispersion} & axis ratio\\
$i$  & $L_i$ [counts] & $I_{0,i}$ [$L_\odot$/pc$^2$] & $\sigma_i$ [$''$] & $\sigma_i$ [kpc] & $q'_i$\\\hline
1  &     9425.96 &      20768.  &  0.051   & 0.103  & 1.00\\
2  &    13173.0 &        3161.2 &  0.178   & 0.358  & 0.76\\
3  &    40235.0 &        1588.2 &  0.503   & 1.008  & 0.58\\
4  &    67755.2 &         502.25&  1.180   & 2.368  & 0.56\\
5  &    203677. &         136.51&  3.891   & 7.805  & 0.57\\\hline
\end{tabular}
\label{tab:MGEF814W}
\end{table*}
%============================================

%--------------------------------------------
\subsubsection{MGE for the inner and outer regions}
%--------------------------------------------

%--------------------------------------------
We fit a MGE to J1331's smooth central region within $\sim 5''$ of the HST/WFPC2/WF3/F814W image (Figure \ref{fig:F814W}). Bright objects close to the bulge (blobs possibly belonging to the background galaxy and parts of the foreground spiral arm) were masked during the fit. J1331's galaxy center, position angle (with respect to North through East), and average apparent ellipticity (see Table \ref{tab:galaxyparameters}) are found from the images weighted first and second moment. The MGE fit splits the image in annuli with the given ellipticity and position angle and sectors of $5^\circ$ width and fits an 5-Gaussian MGE of the form in Equation \eqref{eq:MGEgeneral} convolved with the PSF MGE in Table \ref{tab:PSFMGEF814W} to it. The best-fitting MGE (PSF convolved) is compared to the data in Figure \ref{fig:MGEinnerRegions}, and the corresponding parameters of the intrinsic surface brightness distribution are given in Table \ref{tab:MGEF814W}. The fit is a very good representation of the light distribution in the inner $5''$, but underestimates the light distribution further out.

To get a handle on the light distribution also in the outer parts of J1331 where spiral arms dominate, we first fit a IRAF \citep{1993ASPC...52..173T} \emph{ellipse} model to the F814W image (masking the brightest blobs in the spiral arms and outer regions). Only then we fit a 7-Gaussian MGE to the smooth \emph{ellipse} model. The MGE does not perfectly reproduce the flatness of the \emph{ellipse} model at every radius (see Figure \ref{fig:MGEouterRegions}), but considering the spiral arm dominated outer regions of J1331 it is good enough for an approximate handling of the overall light distribution.
%--------------------------------------------

%--------------------------------------------
\subsubsection{Transformation into physical units}
%--------------------------------------------

%--------------------------------------------
To transform the MGE in units of counts into physical units, we apply a simplified version of the procedure described in \citet{Holtzman}.

The scaling of the drizzled HST/WFC3 images is  $S \equiv 0.05''/\text{pixel width}$ and the total exposure time $T = 1600$ sec. Each Gaussian in the MGE has a total F814W luminosity $L_i$ (in counts) and a central surface brightness (in counts per pixel) of
\begin{equation}
C_{0,i}\text{[counts/pixel]} = \frac{L_i[\text{counts}]}{2\pi \sigma_i[\text{pixel}]^2 q_i}.
\end{equation}
This is then transformed into an $I$-band surface brightness (in $\text{mag}\times(1'')^{-2}$) via
\begin{equation}
\mu_{I,0,i} \simeq -2.5 \log_{10}\left( \frac{C_{0,i}\text{[counts/pixel]}}{T[\text{sec}] \times S[''/\text{pixel width}]^2}\right) + Z + C + A_I, \label{eq:muI_}
\end{equation}
where $Z\simeq21.62~\text{mag}$ is the zero-point from \citet{Holtzman}, updated according to \citet{Dolphin}\footnote{We used the updated zero points by Andrew E. Dolphin, "Zero Points relative to Holtzman et al. (1995)" from \url{http://americano.dolphinsim.com/wfpc2_calib/2008_07_19.html}. The data was retrieved on September 20th, 2013.}, for the photometric system of the HST/WFPC2 camera and the F814W filter, plus a correction for the difference in gain between calibration and observation. $C= 0.1~\text{mag}$ corrects for the finite aperture of the WFPC2; and $A_I =0.015~\text{mag}$ is the extinction in the (Landolt) $I$-band towards J1331, according to the NASA/IPAC Extragalactic Database (NED)\footnote{The NASA/IPAC Extragalactic Database (NED, \url{https://ned.ipac.caltech.edu/}) is operated by the Jet Propulsion Laboratory, California Institute of Technology, under contract with the National Aeronautics and Space Administration. The data for J1331 (SDSS J133140.33+362811.9) was retrieved in October 2013.}. The colour-dependent correction between the F814W filter and the $I$-band of the UBVRI photometric system is  small \citep{Holtzman} and we neglect it therefore. The last step is to transform the surface brightness $\mu_{I,0,i}$ (in mag) to the $I$-band surface density $I_{0,i}$ (in $L_\odot$/pc$^2$) of the Gaussian, i.e.,
\begin{eqnarray}
I_{0,i}[L_\odot \text{pc}^{-2}] = \left( 64800/\pi\right)^2 \left(1+z_{\rm d} \right)^4 10^{0.4\left(M_{\odot,I}-\mu_{I,0,i} \right)},
\end{eqnarray}
where the term with $z_{\rm d}$ accounts for redshift dimming, and $M_{\odot,I}=4.08~\text{mag}$ is the Sun's absolute $I$-band magnitude \citep{1998gaas.book.....B}. The luminosity $L_i$[counts] and the corresponding surface brightness density $I_{0,i} [L_\odot \text{pc}^{-2}]$ of each Gaussian are given in Table \ref{tab:MGEF814W}.
%--------------------------------------------

%--------------------------------------------
\subsubsection{Inclination, total luminosity, and effective radius} \label{sec:incl_Ltot_Reff}
%--------------------------------------------

%--------------------------------------------
To estimate the inclination of J1331 with respect to the observer, we use the observed axis ratio of the flattest ellipse in the IRAF \emph{ellipse} model for J1331, which is $q'=0.42$. This is similar to the disk axis ratio of $q' = 0.40$ found by \citet{SWELLSI}. If a typical thickness of an oblate disk is around $q_0 \sim 0.2$ \citep{1958MeLu2.136....1H}, the inclination follows from 
\begin{equation}
\cos^2 \iota = \frac{q'^2 - q_0^2}{1 - q_0^2}
\end{equation}
and a correction of $+3^\circ$ \citep{1988ngc..book.....T}. Our estimate for the inclination is therefore $\iota \approx 70^\circ$. Given this inclination, the two-dimensional MGE models can be deprojected into three dimensions (see Section \ref{sec:MGE_theo_deprojection}). We also assume the inclination angle to be known and fixed to this value in the dynamical modelling in Sections \ref{sec:results_JAM_SB} and \ref{sec:results_JAM_NFW}.

J1331's total $I$-band luminosity is determined by summing up the luminosity contributions of all the MGE Gaussians for the outer regions (shown as grey lines in Figure \ref{fig:MGEouterRegions}). We find $L_\text{I,tot} \simeq 5.6 \times 10^{10} ~L_\odot$. This corresponds to an apparent magnitude of $m_I = 15.77~\text{mag}$. We determine the circularized effective radius $R_\text{eff}$ of J1331 from the definition $L(<R_\text{eff}) \equiv \frac 12 L_\text{tot}$ and the growth curve $L(<R')$ from the MGE model of the outer regions, where $R'$ is the projected radius on the sky. We find the effective radius to be $R_\text{eff} \simeq 2.6'' \hat{=} 5.2~\text{kpc}$.  

All galaxy parameters are summarized in Table \ref{tab:galaxyparameters}.
%--------------------------------------------

%============================================
\begin{table}
\centering
\caption{Positions of the lensing images (A-D) and the galaxy center (G) in Figure \ref{fig:lens_just_imgpos}. The image positions were determined from the lens-MGE subtracted F450W image for J1331 by \citet{SWELLSIII} (their Figure 4), rotated to the $(x',y')$ coordinate system in Figure \ref{fig:lens_just_imgpos}. The pixel scale is 1 pixel = $0.05''$ and the error of the image positions A-D is $\pm$ 1 pixel.}
\begin{tabular}{r|rrrr|c}
\hline
  & A & B & C & D & G\\\hline
$x_i'$ [pixel] & 12.1 & -8.5 & 21.7 & -3.3 & 0.5 $\pm \sqrt{2}$ \\
$y_i'$ [pixel] & 16.6 & -10.4 & -0.5 & 19.2 & 0.5 $\pm \sqrt{2}$ \\
\hline
\end{tabular}
\label{tab:lenspos}
\end{table}
%============================================

%============================================
\begin{figure}
\centering
  \includegraphics[width=\linewidth]{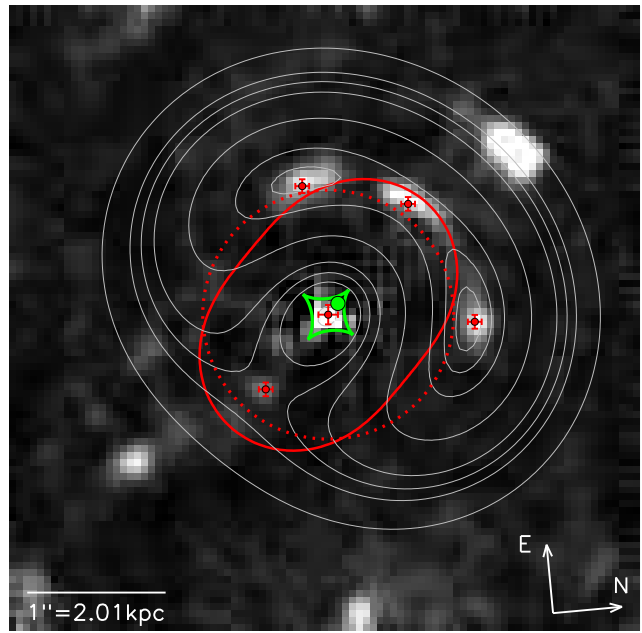}
\caption{Lensing model (in Table \ref{tab:bestfitlensmodel}) found as best fit to the lensing image positions. In the background we show the central region of J1331 in the F450W filter, subtracted by an IRAF \emph{ellipse} model of the F450W surface brightness and smoothed to remove noise smaller than the PSF. The brightness peaks of the four lensing images and the galaxy center (Table \ref{tab:lenspos}) are marked as red dots. For the best-fitting lens model we show the Einstein radius (red dotted circle) and the critical curve (red solid line), which are located in the lens plane. We also show the caustic (green solid line) corresponding to the critical curve and the best-fitting source position (green dot), which are located in the source plane. For $\alpha=1$, the critical curve is a contour of constant surface density of the mass model. The grey  lines show contours of the time delay surface given by Equation (63) in \citet{1996astro.ph..6001N}. Not only the position of the extrema, but also their shape is consistent with the observed, extended images, even though we did not use information about the image shape in the analysis.}
\label{fig:bestfitlensmodel}
\end{figure}
%============================================

%--------------------------------------------
\subsection{Mass distribution from lensing} \label{sec:results_lensing}
%--------------------------------------------

%--------------------------------------------
In this section we use the gravitational lensing formalism summarized in Section \ref{sec:lensing_theo} to fit a scale-free galaxy mass model to the positions of the lensing images observed in J1331's central region.
%--------------------------------------------

%============================================
\begin{table*}
\centering
\caption{Best-fitting lens model (first column) found from the peak image positions in Table \ref{tab:lenspos} following the procedure described in Section \ref{sec:lensing_theo} and assuming a flat rotation curve ($\alpha = 1$). The second column gives the corresponding best-fitting mean and standard deviation derived from Monte Carlo sampling of the Gaussian uncertainties around the image positions, with the relative error given in parentheses.}
\begin{tabular}{llrrclr}
\hline
 &  & \multicolumn{1}{c}{lens model for} &\multicolumn{4}{c}{lens model from Monte Carlo sampling  } \\
 &  & \multicolumn{1}{c}{peak image positions}  & \multicolumn{4}{c}{of image position uncertainties}  \\ \hline
Einstein radius      & $R_\text{ein}$ [$''$]             & $0.907$ & $0.91$  & $\pm$ & $     0.02$ & ($2\%$)\\
Einstein mass        & $M_\text{ein}$ [$10^{10} \text{M}_\odot$]  & $7.72$  & $7.8 $  & $\pm$ & $      0.3$ & ($4\%$) \\
Critical mass        & $M_\text{crit}$ [$10^{10} \text{M}_\odot$] & $7.87$  & $7.9$   & $\pm$ & $      0.3$ & ($4\%$)\\
Source position      & $x_{\rm s}'$ [$''$]                      & $0.095$ & $0.09 $ & $\pm$ & $     0.03$ & ($28\%$)\\
                     & $y_{\rm s}'$ [$''$]                     & $0.107$ & $0.10 $ & $\pm$ & $     0.03$ & ($27\%$)\\
Fourier coefficients & $a_0$                               & $1.814$ & $1.82 $ & $\pm$ & $   0.04$ & (2\%)\\
                     & $a_2$                               & $0.012$ & $ 0.011 $ & $\pm$ & $    0.004$ & (35\%)\\
                     & $b_2$                               & $-0.057$ & $-0.06 $  & $\pm$ & $  0.01$ & (25\%)\\
                     & $a_3$                               & $-0.0001$& $0.0000 $ & $\pm$ & $   0.0006$ & \\
                     & $b_3$                               & $-0.0002$&$0.000 $   & $\pm$ & $  0.001$ & \\\hline
\end{tabular}  
\label{tab:bestfitlensmodel} 
\end{table*}
%============================================

%============================================
\begin{figure*}
\centering
\begin{subfigure}{.3\textwidth}
  \centering
  \includegraphics[width=.9\linewidth]{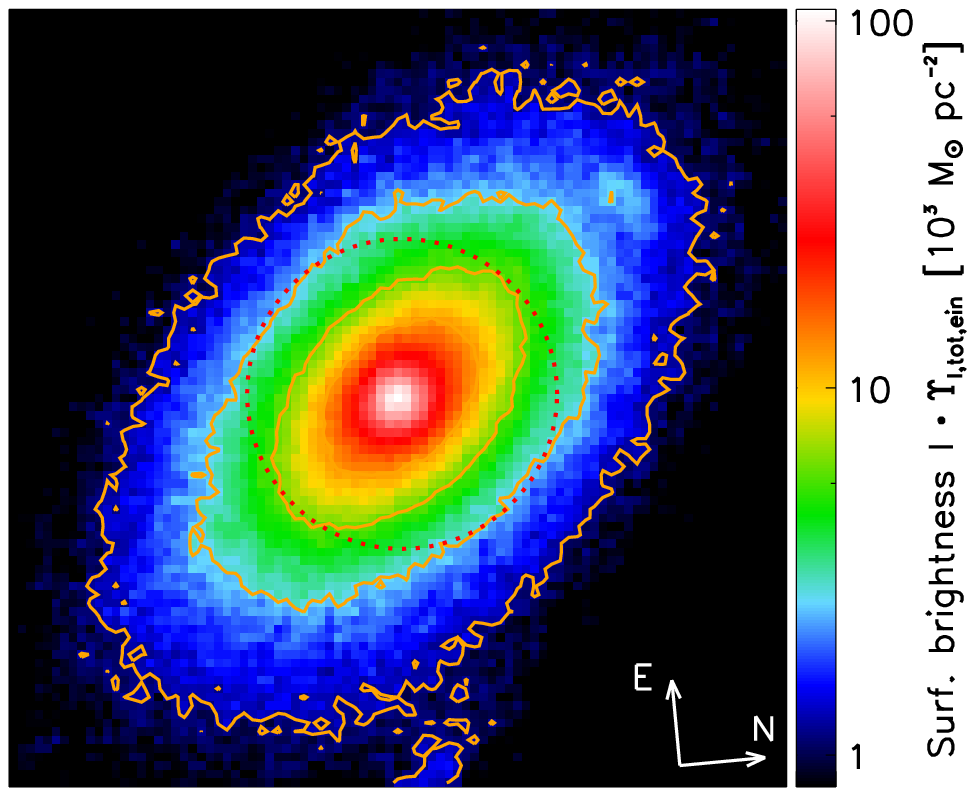}
  \caption{Observed light distribution.}
  \label{fig:lenscomparelight}
\end{subfigure}%
\begin{subfigure}{.3\textwidth}
  \centering
  \includegraphics[width=.9\linewidth]{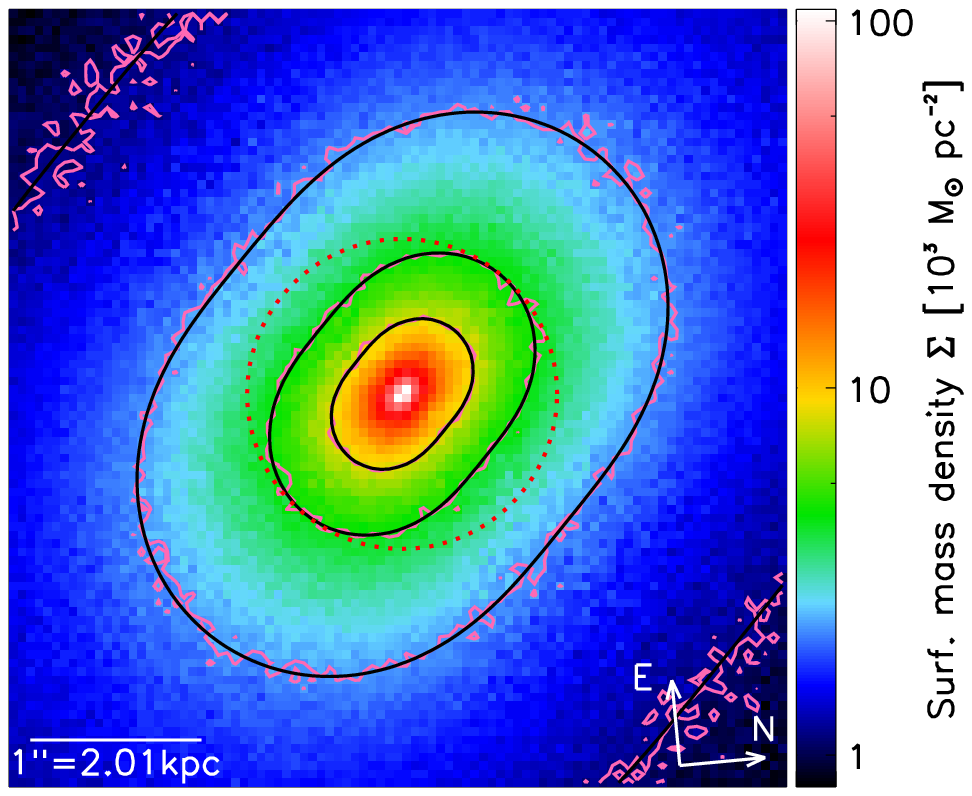}
  \caption{Lens model mass distribution.}
  \label{fig:lenscomparemass}
\end{subfigure}
\begin{subfigure}{.3\textwidth}
  \centering
  \includegraphics[width=.9\linewidth]{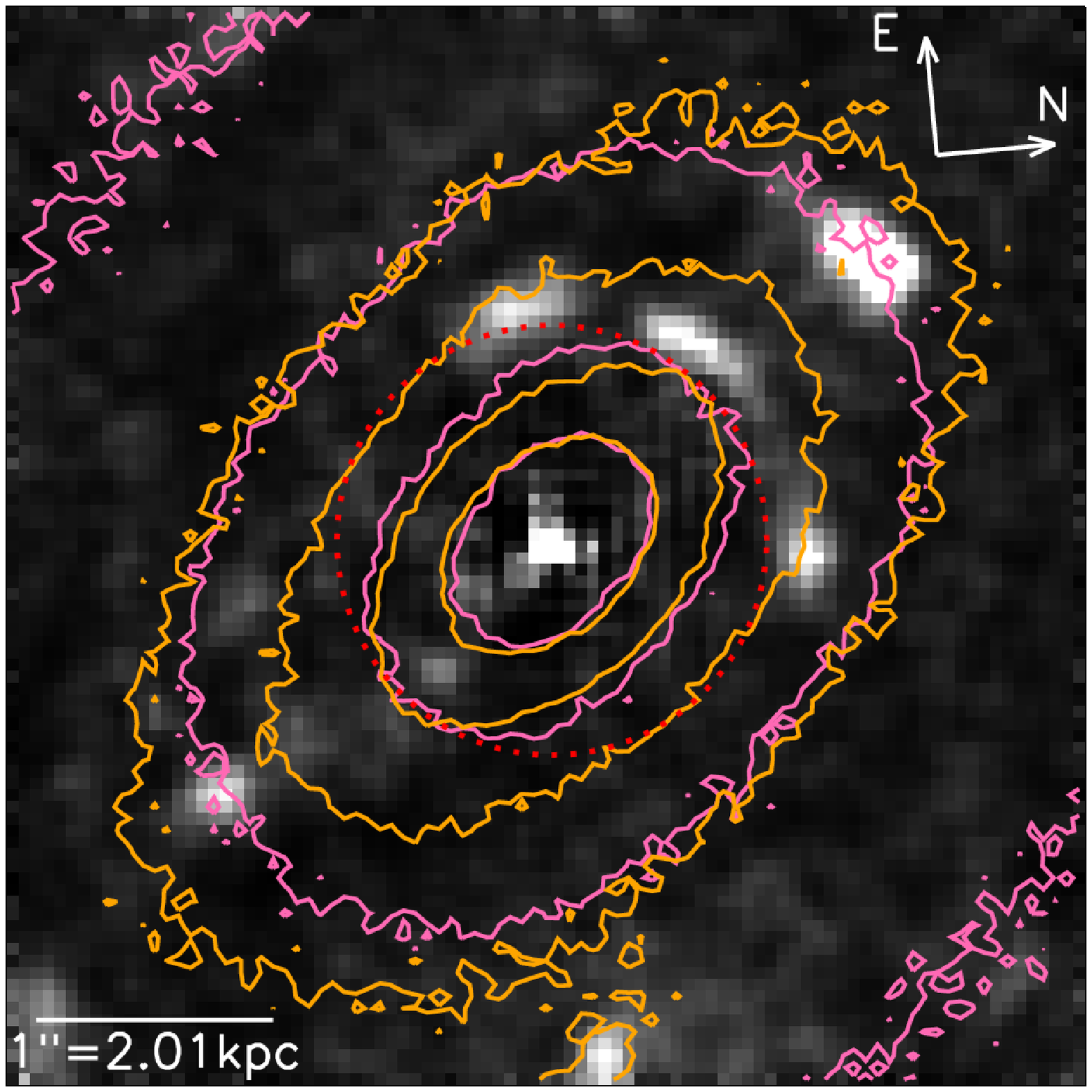}
  \caption{Comparison of mass and light.}
  \label{fig:lenscompareboth}
\end{subfigure}
\caption{Comparison of the observed F814W/$I$-band surface brightness distribution (panel (a) and orange contours) and derived mass distribution from lensing constraints (panel (b) and pink contours). To allow for a qualitative comparison of the contours in panel (c), the light distribution was turned into a mass distribution by multiplication with the total mass-to-light ratio inside the Einstein radius, $\Upsilon_\text{I,tot}^\text{ein} = 5.56 \Upsilon_{I,\odot}$. The Einstein radius is overplotted as red dotted circle. The uncertainties in the mass model in the second column of Table \ref{tab:bestfitlensmodel} were translated into random Monte Carlo noise in the mass contours. The smooth black contours correspond to the best-fitting model in the first column of Table \ref{tab:bestfitlensmodel}. We find that the lens mass model is more roundish than the light distribution. The background in panel (c) shows again the surface brightness subtracted center of the galaxy to make the lensing images visible.}
\label{fig:lenslightcompareALL}
\end{figure*}
%============================================

%--------------------------------------------
\subsubsection{Image positions}
%--------------------------------------------

%--------------------------------------------
We determine the positions of the lensing images by first subtracting a smooth model for the galaxy's surface brightness from the original image. As models we use MGE fits and IRAF \emph{ellipse} fits to J1331 in each the F450W and F814W filter. The lensing images become visible in the residuals (see Figure \ref{fig:lens_just_imgpos}). Because the images are extended, we use the position of the brightest pixel in each of the images. In addition, we consider the F450W-MGE subtracted residuals from \citet{SWELLSIII}. The lensing positions, as determined from the latter, are given in Table \ref{tab:lenspos}. The scatter of lensing positions, as determined from subtracting different brightness models from the galaxy in different filters, gives an error of $\pm 1$ pixel on the image positions. To the galaxy center, which we assume to be the surface brightness peak in the F450W image, we apply an error of $\pm \sqrt{2}$ pixel.

Eight image position coordinates allow us to fit a lens mass model with only $<8$ free parameters. We therefore do not fit Fourier components $(a_k,b_k)$ with $k > 3$ in the lens mass model in Equations \eqref{eq:scalefreemodel}-\eqref{eq:Fourieransatz}.

Even though the constraint from the image positions on $\alpha$ is very weak, we were able to show that the image positions in Table \ref{tab:lenspos} are consistent with a model with flat rotation curve. In the following we therefore set $\alpha=1$.
%--------------------------------------------

%--------------------------------------------
\subsubsection{Best-fitting lens model} \label{sec:results_lensing_bestfit}
%--------------------------------------------

%--------------------------------------------
We fit the lens mass model to the image positions in Table \ref{tab:lenspos} by minimizing $\chi^2 = \chi_\text{lens}^2 + \chi_\text{shape}^2$ (see Equations \eqref{eq:chi2lens}-\eqref{eq:chi2total}). The best-fitting parameters are given in the first column of Table \ref{tab:bestfitlensmodel}. Figure \ref{fig:bestfitlensmodel} shows the corresponding critical curve, caustic and Einstein radius, and the best-fitting source position. In this case, where $\alpha=1$, the (tangential) critical curve is also an equidensity contour of the galaxy model \citep{EvansWitt}, which appears to have an elliptical mass distribution. The source is located close to a cusp of the diamond-shaped caustic: a lensing configuration for which we indeed expect four images. Figure \ref{fig:bestfitlensmodel} also shows the (smoothed) residuals from the F450W image subtracted by an IRAF \emph{ellipse} brightness model and the contours of the best-fitting model's time delay surface (see \S 3.3.1 in \citealt{1996astro.ph..6001N}).  Fermat's principle states that the image positions are observed at the extrema of the time delay surface. And even though we did not include any information about the shape of the lensing images in the fit, it is consistent with the predicted distortion for an extended source.

To estimate how the uncertainties in the determination of image positions and galaxy center affect the results, we sample random positions from two-dimensional Normal distributions with peaks and standard deviations according to Table \ref{tab:lenspos}. Model fits to many sampled image positions lead to probability distributions for the best-fitting shape parameters and Einstein quantities; peak and standard deviations are given in the second column of Table \ref{tab:bestfitlensmodel}. We constrain the Einstein radius to within 2\%, $R_\text{ein} = (0.91 \pm 0.02)'' \hat{=}(1.83\pm0.04)~\text{kpc}$, and the projected mass within the critical curve with a relative error of 4\%, $M_\text{crit} =(7.9\pm0.3) \times 10^{10} \text{M}_\odot$. Our measurement of $R_\text{ein}$ is consistent with that from \citet{SWELLSIII}, $R_\text{ein,SWELLS} = (0.96 \pm 0.04)''$ (which used a singular isothermal ellipsoid as lens mass model and the intermediate-axis convention of the critical curve as the Einstein radius). The relative difference between our critical mass and that of \citet{SWELLSIII}, $M_\text{crit,SWELLS} =(8.86\pm0.61) \times 10^{10} \text{M}_\odot$, is 13\%.
%--------------------------------------------

%============================================
\begin{figure}
  \centering
  \includegraphics[height=6cm]{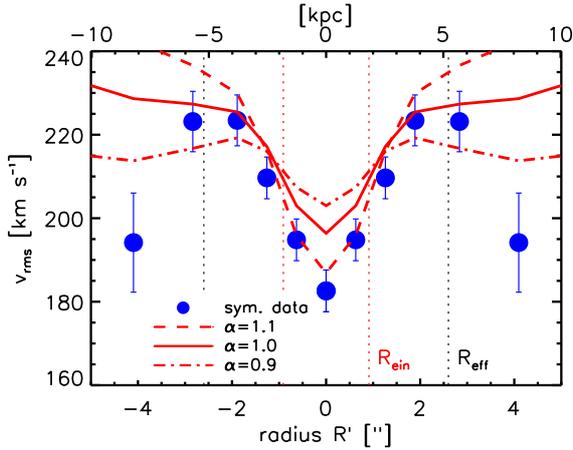}
  \caption{Comparison (not a fit!) of the symmetrized stellar $v_\text{rms}$ data of J1331 (blue dots) with JAM models generated from mass distributions which were independently derived from lensing constraints in Section \ref{sec:results_lensing} (red lines). The red solid line corresponds to the lens model for a flat rotation curve ($\alpha = 1$) in Table \ref{tab:bestfitlensmodel}; the red dashed and dash-dotted lines are the best-fitting lens models found analogously from the image positions, but for a fixed rotation curve slope of $\alpha = 1.1$ and $0.9$, respectively. For the JAM modelling, best-fitting MGEs to the lens mass models were used, as well as the observed surface brightness MGE in Table \ref{tab:MGEF814W}, assuming velocity isotropy $\beta_z = 0$ and an inclination of $\iota = 70^\circ$. The most reliable constraints are around the Einstein radius (vertical red dotted line); outside of it the model is just an extrapolation.}
  \label{fig:JAM_modelL}
\end{figure}
%============================================

%--------------------------------------------
\subsubsection{Comparison with the light distribution} \label{sec:results_lensing_compare}
%--------------------------------------------

%--------------------------------------------
The surface mass distribution as predicted by the best-fitting lens model (Table \ref{tab:bestfitlensmodel}) is shown in Figure \ref{fig:lenscomparemass}. We visualize the effect of the Fourier shape parameter uncertainties by introducing random noise to create a mock observation. From the mock image's second moment we find an average axis ratio for the lens mass model of $q_\text{lens} \simeq 0.695$, which is consistent with the one found by \citet{SWELLSIII}, $q_\text{lens,SWELLSIII} = 0.67 \pm 0.09$, while the light's average axis ratio is $q' = 0.598$ (see Table \ref{tab:galaxyparameters}).

We estimate the total (projected) mass-to-light ratio within the Einstein radius $\Upsilon_\text{I,tot}^\text{ein} \equiv M_\text{ein} / L_\text{I,ein}$. For this, we integrate the MGE in Table \ref{tab:MGEF814W} to get the total $I$-band luminosity within the Einstein radius $L_\text{I,ein} = 1.40 \times 10^{10} L_\odot$. The corresponding Einstein mass-to-light ratio is therefore $\Upsilon_\text{I,tot}^\text{ein} = 5.56 \Upsilon_{I,\odot}$. This is consistent with or slightly larger than the stellar mass-to-light ratio assuming a Salpeter initial mass function (IMF) $\Upsilon_\text{I,*}^\text{sal} = 4.7 \pm 1.2$ according to \citet{SWELLSI} and Table \ref{tab:previousresults} (see also discussion in Section \ref{sec:MLdiscussion}).

We use $\Upsilon_\text{I,tot}^\text{ein}$ to transform the observed surface brightness in the F814W filter into a surface mass density (Figure \ref{fig:lenscomparelight}). Figure \ref{fig:lenscompareboth} then compares equidensity contours of both the predicted lens mass distribution and the observed surface brightness times $\Upsilon_\text{I,tot}^\text{ein}$.

Figure \ref{fig:lenslightcompareALL} leads to the following three findings: (i)The mass predicted from lensing and the observed light distribution are oriented in the same direction (i.e., have the same position angle). (ii) Within and around the Einstein radius, mass and light distribution have a similar elliptical shape, while further out the mass distribution is slightly rounder. (iii) The light distribution drops faster than the mass with increasing radius (which is---at least partly---because of the assumption of a flat rotation curve). However, the mass distribution is only constrained around the Einstein radius and otherwise is an extrapolation. 
%--------------------------------------------

%============================================
\begin{figure}
  \centering
  \includegraphics[height=6cm]{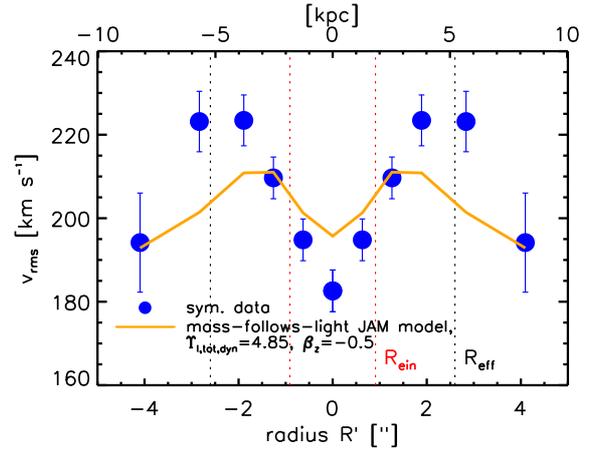}
  \caption{Comparison of the symmetrized $v_\text{rms}$ data of J1331 (blue dots) with a best-fitting dynamical JAM model (orange line) assuming ``mass-follows-light'' and with two free parameters: $\Upsilon_\text{I,tot}^\text{dyn}$, the total constant $I$-band mass-to-light ratio which converts the observed surface brightness in Table \ref{tab:MGEF814W} into a mass distribution, and the velocity anisotropy parameter $\beta_z$. The ``best'' fit is $\Upsilon_\text{I,tot}^\text{dyn} = 4.8 \pm 0.1$ and $\beta_z = -0.5$, where the latter is however pegged at the lower limit of the allowed value range. This is obviously not a good model.}
  \label{fig:JAM_modelA2}
\end{figure}
%============================================

%--------------------------------------------
\subsection{JAM based on surface brightness} \label{sec:results_JAM_SB}
%--------------------------------------------

%--------------------------------------------
In this section we create dynamical models for J1331 following the procedure in Sections \ref{sec:model_JAM_JAM}-\ref{sec:model_JAM_compare}. We use the deprojected surface brightness MGE from Table \ref{tab:MGEF814W} for the tracer distribution $\nu(R,z)$ and to generate mass models $\rho(R,z)$ (assuming the galaxy's inclination angle to be $\iota=70^\circ$; see Table \ref{tab:galaxyparameters} and Section \ref{sec:incl_Ltot_Reff}). The only exception is the first test, where the mass model comes from lensing constraints.
%--------------------------------------------

%============================================
\begin{figure*}
\centering
\begin{subfigure}{.5\textwidth}
  \centering
  \includegraphics[height=6cm]{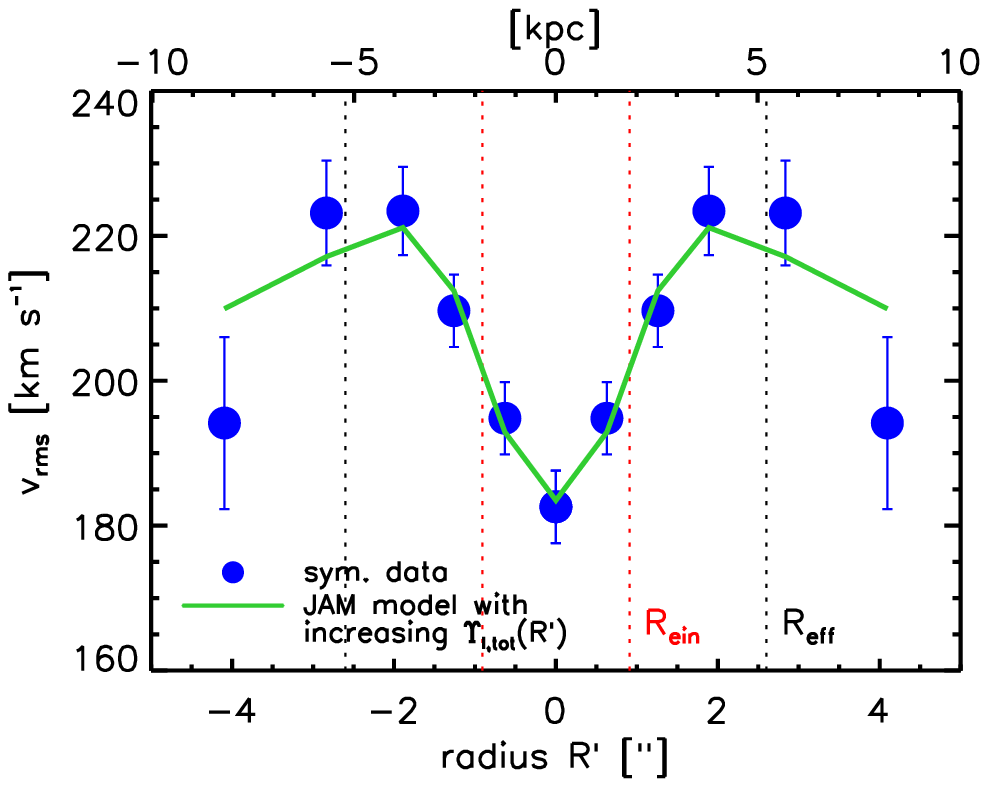}
  \caption{Comparison of $v_\text{rms}$ data and best-fitting model.}
  \label{fig:JAM_modelG}
\end{subfigure}%
\begin{subfigure}{.5\textwidth}
  \centering
  \includegraphics[height=6cm]{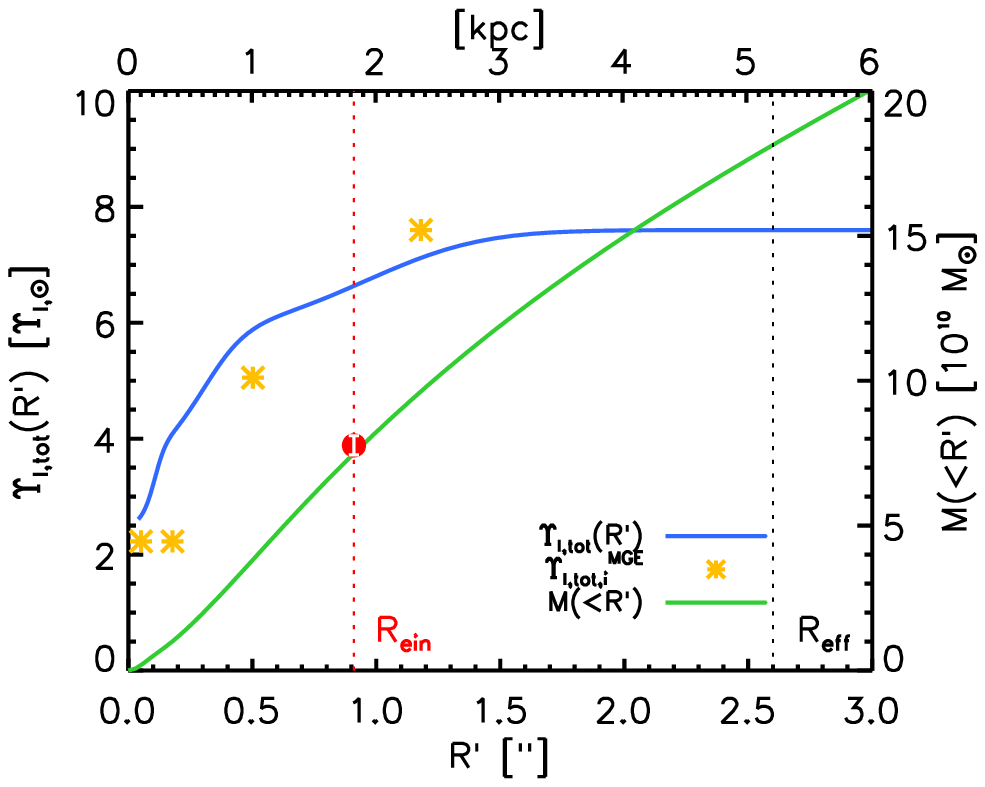}
  \caption{Projected local mass-to-light ratio profile and enclosed mass.}
  \label{fig:enclMass_modelG}
\end{subfigure}
\caption{JAM model found by fitting a galaxy model with increasing total mass-to-light ratio profile $\Upsilon_\text{I,tot}(R')$ to the symmetrized $v_\text{rms}$ data (blue dots). The mass distribution is generated by assigning a different mass-to-light ratio $\Upsilon_{\text{I,tot,}i}$ to each Gaussian $i$ in the light distribution MGE in Table \ref{tab:MGEF814W}. The $\Upsilon_{\text{I,tot,}i}$ were treated as free fit parameters. \emph{Panel (a):} Comparison between the stellar $v_\text{rms}$ data (blue dots) and the best-fitting model (green line). \emph{Panel (b):} Projected total mass-to-light profile $\Upsilon_\text{I,tot}(R')$ along the major axis (blue line, left axis) of the best-fitting model. The best-fitting mass-to-light ratios $\Upsilon_{\text{I,tot,}i}$ are plotted against $\sigma_i$ for each MGE Gaussian ($i<5$, yellow stars). (The Gaussians $i=4$ and $i=5$ have the same $\Upsilon_{\text{I,tot,}i}$.) Shown is also the enclosed mass inside the projected radius $R'$ on the sky, $M(<R')$ (green line, right axis). The enclosed mass curve is overplotted with the independent finding for the Einstein mass $\pm 4 \%$ in Table \ref{tab:bestfitlensmodel} (red dot) at the Einstein radius (red dotted line); the agreement is very good.}
\label{fig:modelG}
\end{figure*}
%============================================

%--------------------------------------------
\subsubsection{JAM with lens mass model}
%--------------------------------------------

%--------------------------------------------
We make an independent prediction for the $v_\text{rms}$ curve by evaluating the JAM equations (with $\beta_z = 0$) for the lens mass model in Table \ref{tab:bestfitlensmodel} ($\alpha = 1$, flat rotation curve). In addition, we also calculate a  $v_\text{rms}$ curve for two lens models which were found analogously, but assumed a slightly rising (falling) rotation curve slope of $\alpha=1.1$ ($\alpha=0.9$). The predictions are compared with the data in Figure \ref{fig:JAM_modelL}. While the most reliable constraint is around $R_\text{ein}$, the agreement within $R' \sim 3''$ is still striking: The $\alpha > 1$ model recreates the observed central dip, while the $\alpha = 1$ model fits the wings around $R_\text{eff}$. This is in concordance with observations in other galaxies. For $R'> R_\text{eff}$ we would expect $\alpha<1$; and the lensing model for $\alpha=0.9$ has indeed a slightly dropping $v_\text{rms}$ around $R_\text{eff}$ like the data. A definite comparison in this regime is however difficult as the lens models are just extrapolations outside of $R_\text{ein}$. While our lensing model assumes $\alpha(R')=\text{const}$, Figure \ref{fig:JAM_modelL} suggests that a model with variable $\alpha(R')$ could fit even better. Overall the lensing model is in very good agreement with the $v_\text{rms}$ data within $R' \sim 3''$, even though it was derived completely independently.
%--------------------------------------------

%--------------------------------------------
\subsubsection{JAM with "mass-follows-light" and velocity anisotropy} \label{sec:results_JAM_SB_MfL}
%--------------------------------------------

%--------------------------------------------
The first JAM model that we fit to the observed $v_\text{rms}$ within $R'<5''$ is a mass-follows-light model. Mass-follows-light models are often used in dynamical JAM modelling (e.g., \citealt{GlennEC,Cap06}) and generate a mass distribution by multiplying the intrinsic light distribution $\nu(R,z)$ by a constant total mass-to-light ratio  $\Upsilon_\text{I,tot}^\text{dyn}$. This assumes that the DM is always a constant fraction of the total matter distribution within the region covered by the kinematics. This simplified mass model sometimes gives good representations of the inner parts of galaxies where the stellar component dominates.

We also allow for an overall constant but non-zero velocity anisotropy $\beta_z$ in the model. The model parameters ($\Upsilon_\text{I,tot}^\text{dyn},\beta_z$) that fit the $v_\text{rms}$ data best are found using a $\chi^2$-fit and are demonstrated in Figure \ref{fig:JAM_modelA2}. 

For $\beta_z$ we imposed the fitting limits $\beta_z \in [-0.5,+0.5]$. While the outer parts of galaxies often show radially biased velocity anisotropy up to $\sim 0.5$ (from dynamical modelling of observed elliptical galaxies, e.g., \citet{Kronawitter2000}) and cosmological simulations (e.g., \citealt{2004MNRAS.352..535D,2001ApJ...557..533F}), the centers of galaxies are near-isotropic or have negative velocity anisotropy \citep{2003ApJ...583...92G}. Only in extreme models (e.g., around in-spiralling supermassive black holes, e.g., \citealt{1997NewA....2..533Q}) velocity anisotropies as low as $\sim -1$ have been found.

The best fit in Figure \ref{fig:JAM_modelA2} however strives to very negative $\beta_z$ to be able to reproduce the deep central dip in the  $v_\text{rms}$ curve.\footnote{Without limiting the fitting range, the best fit would be a unrealistically low $\beta_z \sim -2$.} But $\beta_z = -0.5$ is not even a remotely agreeable fit and lower anisotropies are not to be expected or realistic. We also tested radial profiles for $\beta_z(R)$ of the form proposed by \citet{BaesVanHese}, which was however equally unable to reproduce the data. We conclude that this is due to the well-known degeneracy between anisotropy and mass profile and the mass-follows-light model is \emph{not} a good representation of the mass distribution in J1331's inner regions.
%--------------------------------------------

%--------------------------------------------
\subsubsection{JAM with increasing mass-to-light ratio} \label{sec:results_JAM_SB_gradient}
%--------------------------------------------

%--------------------------------------------
In Section \ref{sec:results_lensing_compare} we found from lensing constraints that the light distribution might drop faster with radius than the mass distribution. This could correspond to a radially increasing total mass-to-light ratio. As velocity anisotropy alone cannot explain the observed kinematics in a simple mass-follows-light model, we now allow for a mass-to-light ratio gradient in the JAM modelling. We therefore generate a mass model from the light distribution in Table \ref{tab:MGEF814W} by assigning each of the five Gaussians in the MGE its own total mass-to-light ratio $\Upsilon_{\text{I,tot,}i}$ and replace the total luminosity in Equation \eqref{eq:centralItotalL}, $L_i$, with the Gaussians total mass $M_i = \Upsilon_{\text{I,tot,}i} \times L_i$. We treat the five $\Upsilon_{\text{I,tot,}i}$ as free fit parameters and only require that $\Upsilon_{\text{I,tot,}j} \geq \Upsilon_{\text{I,tot,}i}$ when the corresponding $\sigma_j \geq \sigma_i$ to ensure that the overall mass-to-light ratio is increasing with radius.

Figure \ref{fig:enclMass_modelG} shows the best-fitting (projected local) mass-to-light ratio profile, which rises from $\Upsilon_\text{I,tot} = 2.53$ in the center and approaches a value of $\Upsilon_\text{I,tot} = 7.60$ outside of the fitted region at $R'\gtrsim 3''$. The central mass-to-light ratio is in agreement with the mass-to-light ratio $\Upsilon_\text{I,*}^\text{chab} = 2.5 \pm 0.6$, given in Table \ref{tab:previousresults} based on the results of \citet{SWELLSI} assuming a stellar population with a \citet{Chabrier2003} IMF (see also Section \ref{sec:MLdiscussion}). When assuming that galaxy bulges are in general older and redder in the center, i.e., $\Upsilon_\text{I,*}$ is more likely to drop with radius than rise, the strong increase of $\Upsilon_\text{I,tot}(R')$ might be due to a strong contribution of DM in J1331.

Figure \ref{fig:JAM_modelG} shows that the best-fitting model nicely reproduces the central dip in the $v_\text{rms}$ curve, even though it has difficulties fitting the drop around $R' \sim 4''$. The latter is because we only allowed the $\Upsilon_\text{I,tot}(R')$ to rise. A slight drop could be expected when the reddish bulge turns into the bluish disk and the contribution of the stellar component becomes less due to a lower $\Upsilon_\text{I,*}$ for younger and bluer populations. Corresponding fitting attempts with the matter model at hand were however unsuccessful: Leaving all five $\Upsilon_{\text{I,tot,}i}$ free lead to over-fitting in the central regions of J1331, while not allowing for sufficient flexibility further out at $R'\gtrsim4''$ to properly model the transition from bulge to disk (because four of the five $\sigma_i$ in the light MGE in Table \ref{tab:MGEF814W} are $< 1.2''$).

For the inner regions the model with rising $\Upsilon(R')$ seems however to be a very good model: In Figure \ref{fig:enclMass_modelG} we overplot the enclosed mass profile with the Einstein mass $M_\text{ein} = (7.77 \pm 0.33) \times 10^{10} \text{M}_\odot$ at the Einstein radius found from lensing in Table \ref{tab:bestfitlensmodel}. The agreement between the Einstein mass and the independently found $M(<R_\text{ein}) = 7.49 \times 10^{10} \text{M}_\odot$ from dynamical modelling is striking.
%--------------------------------------------

%============================================
\begin{figure*}
\centering
\includegraphics[width=0.9\linewidth]{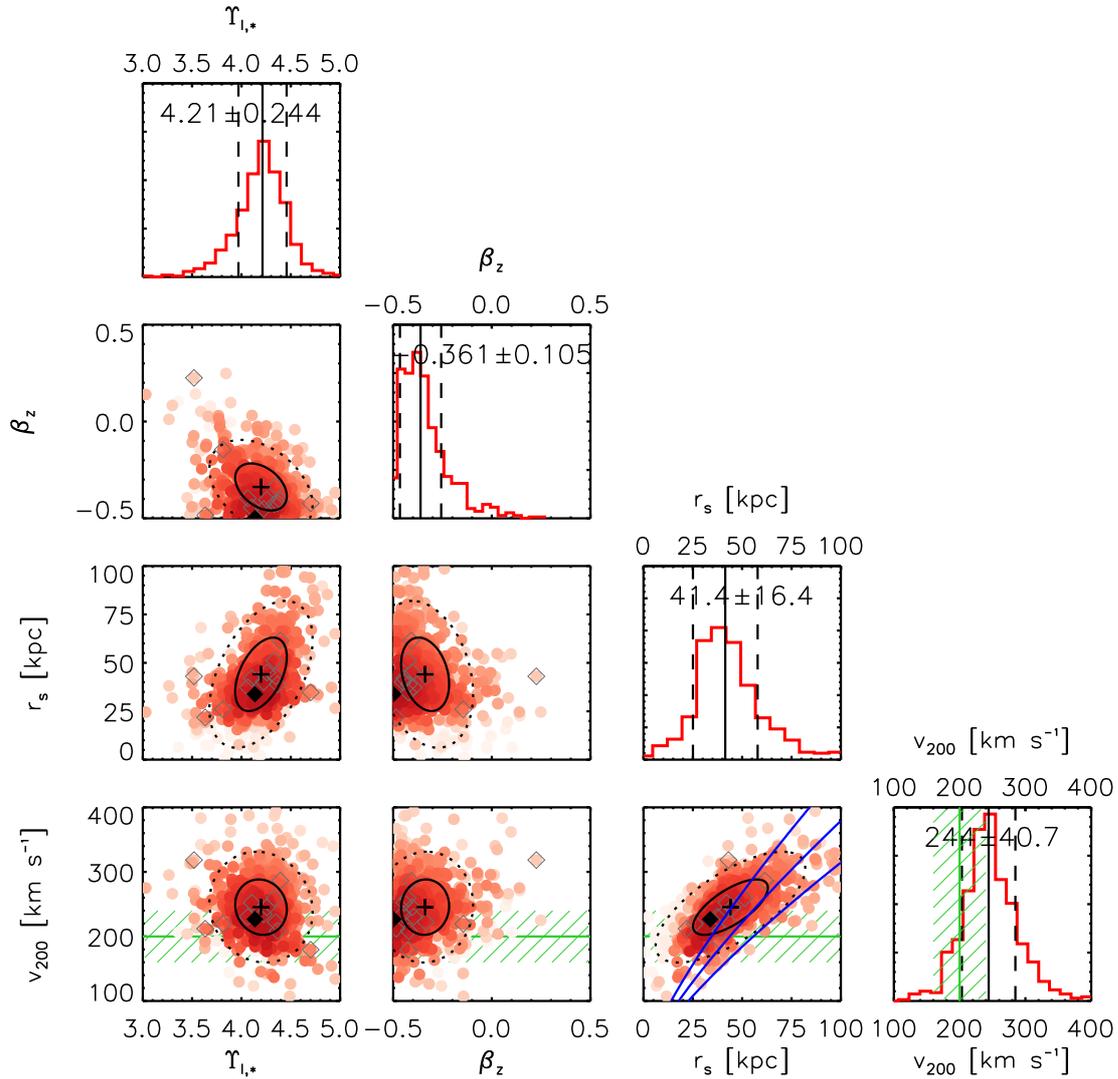}
\caption{Posterior probability distribution (pdf) sampled with a Monte Carlo Markov Chain (MCMC) (red dots and histograms) for a JAM model fitted to the stellar $v_\text{rms}$ data, with NFW halo, parametrized by $r_{\rm s}$ and $v_\text{200}$, a stellar mass distribution generated from the $I$-band MGE in Table \ref{tab:MGEF814W} and a constant stellar mass-to-light ratio $\Upsilon_\text{I,*}$, and constant velocity anisotropy $\beta_z$. Shown are also the priors used for J1331's NFW halo, $\mathscr{N}(200~\text{km s}^{-1},40~\text{km s}^{-1})$ (green, see Equation \eqref{eq:prior_v200}) and the concentration vs. halo mass relation by \citet{Maccio08} from Equation \eqref{eq:Maccio08} in terms of $v_{200}$ vs. $r_{\rm s}$ with $1\sigma$ scatter (blue, see Equation \eqref{eq:prior_c200}). We find slight covariances between $\Upsilon_\text{I,*}$,  $r_{\rm s}$ and $\beta_z$: The smaller the DM contribution in the center (i.e., the larger $\Upsilon_\text{I,*}$), the less concentrated is the halo (i.e., the larger $r_{\rm s}$) and the more velocity anisotropy is needed to reproduce the central $v_\text{rms}$ dip. As the effect of $v_{200}$ is mostly at larger radii, this parameter does not show any covariances, but is also mostly constrained by the prior. The MCMC samples are colour coded according to their probability (darker red for higher probability); the sample point with the highest probability is marked by a black diamond. The black cross is the mean of the distribution and the ellipses are derived from the covariance matrix of the MCMC samples and correspond approximately to $1\sigma$ (black solid ellipse) and $2\sigma$ (black dotted ellipse) confidence. The histograms of the marginalized 1D distributions are overplotted by the mean (black solid lines) and $1\sigma$ error (black dashed lines), whose values are also quoted in the figure and in Table \ref{tab:modelB4_bestfit}. The grey diamonds mark a random sub-selection of 12 samples; the corresponding models are shown in Figure \ref{fig:modelB4_models}.}
\label{fig:modelB4_triangle}
\end{figure*}
%============================================

%============================================
\begin{figure*}
\centering
\begin{subfigure}{.48\textwidth}
  \centering
  \includegraphics[width=0.9\linewidth]{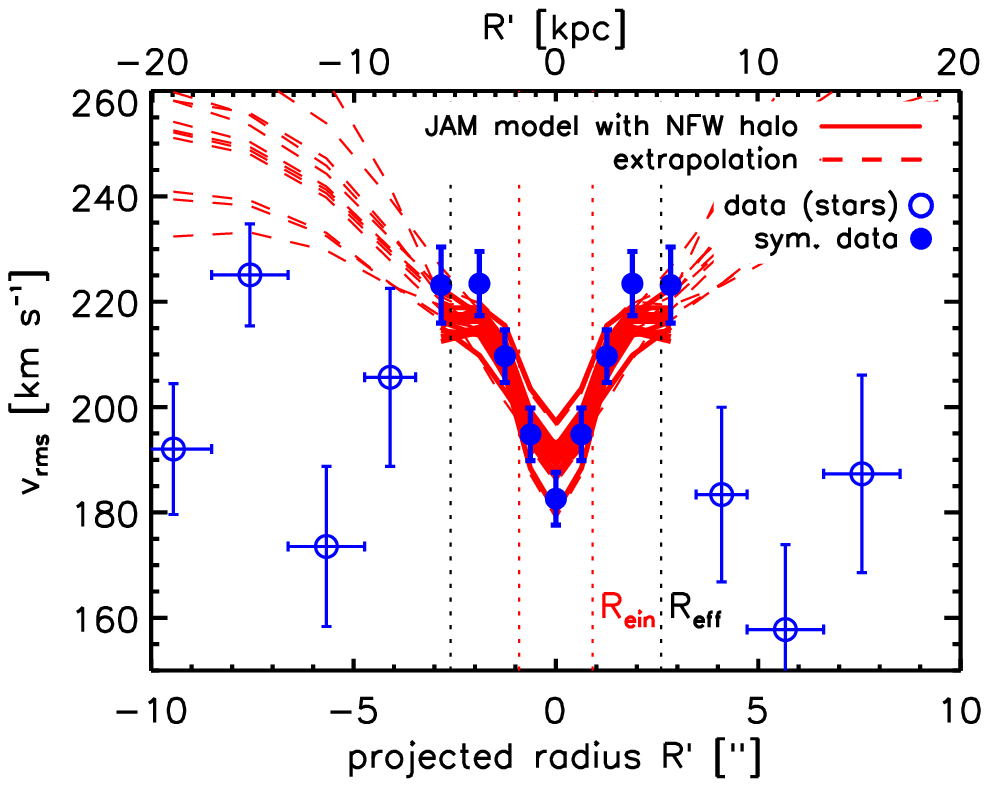}
  \caption{Comparison of the $v_\text{rms}$ data with the best-fitting JAM models.}
  \label{fig:modelB4_vrms}
\end{subfigure}%
\hspace{.02\textwidth}
\begin{subfigure}{.48\textwidth}
  \centering
  \includegraphics[width=0.9\linewidth]{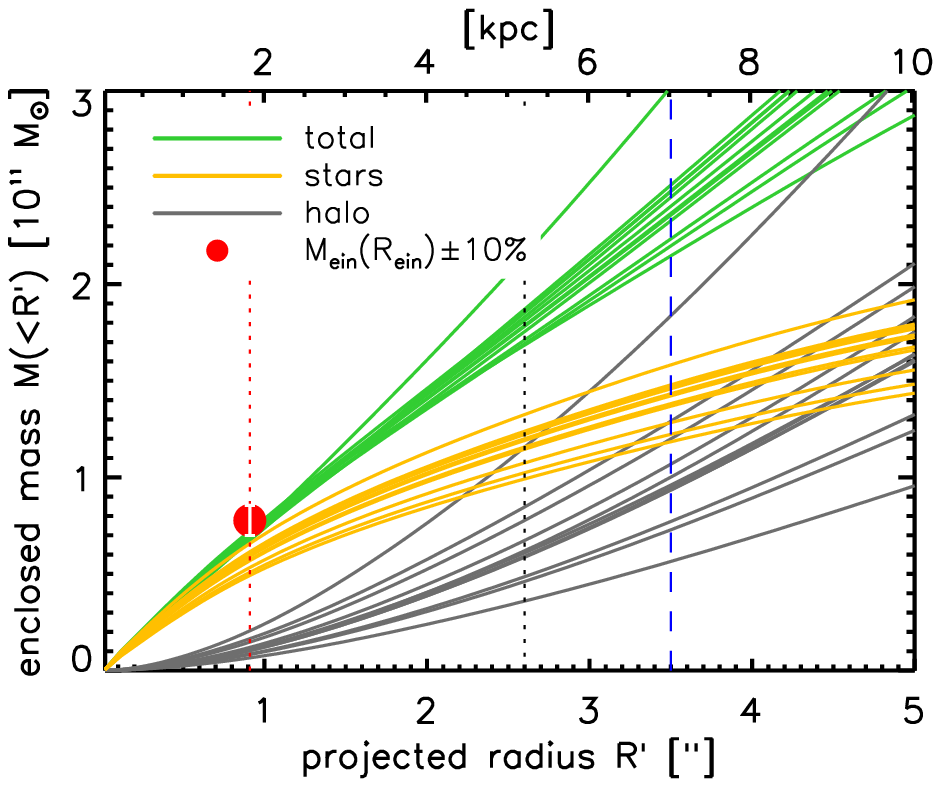}
  \caption{Enclosed mass profiles for the best-fitting models.}
  \label{fig:modelB4_enclMass}
\end{subfigure}
\caption{JAM model including an NFW halo and constant velocity anisotropy fitted to the stellar $v_\text{rms}$ data (blue solid dots) within $\sim3.5''$ and the Einstein mass (red solid dot). The 12 lines shown correspond to 12 models randomly drawn from the pdf (marked as grey diamonds in Figure \ref{fig:modelB4_triangle}) and their range demonstrates how tight the modelling constraints are. The mean and standard deviation of the pdf, i.e., the "best fit" parameters, are given in Table \ref{tab:modelB4_bestfit}. \emph{Panel (a):} At $R'\lesssim 3.5''$ the symmetrized $v_\text{rms}$ data used in the fit (solid blue points) is shown together with the best-fitting $v_\text{rms}$ models (red solid lines). At larger radii the non-symmetrized data (open blue dots) is compared to an extrapolation of the best-fitting models using the surface brightness MGE for the outer regions from Figure \ref{fig:MGEouterRegions} (red dashed lines). \emph{Panel (b):} Shown is the projected enclosed mass profile of the total mass (green), and separately the contribution of the stellar mass (yellow, again generated from the MGE in Table \ref{tab:MGEF814W}) and DM (grey). At the Einstein radius $R_\text{ein}$ (red dotted line) the Einstein mass $M_\text{ein}$ is overplotted with a 10\% error, which was also included in the fit as additional constraint. Overplotted is also the effective half-light radius (black dotted line) and the blue dashed line marks the radius within which the data and model were fitted.}
\label{fig:modelB4_models}
\end{figure*}
%============================================

%============================================
\begin{table*}
\centering
\begin{tabular}{llrcl}
\hline
stellar $I$-band mass-to-light ratio & $\Upsilon_\text{I,*}$ & 4.2 & $\pm$ & 0.2\\
velocity anisotropy & $\beta_z$ & -0.4 & $\pm$ & 0.1 \\
NFW halo scale length & $r_{\rm s}$ [kpc] & 40 & $\pm$ & 20\\
NFW halo virial velocity & $v_{200}$ [$\text{km s}^{-1}$] & 240 & $\pm$ & 40\\
NFW halo concentration & $c_{200}$ & 8 & $\pm$ & 2 \\
NFW halo mass & $M_{200}$ [$10^{12} \text{M}_\odot$] & 5 & $\pm$ & 2\\
\hline
\end{tabular}
\caption{Summary of the best-fitting parameters of the JAM model with NFW halo and constant velocity anisotropy, fitted to the $v_\text{rms}$ data within $\sim 3.5''$ and the Einstein mass $\pm10\%$. These estimates correspond to the mean and standard deviation of the pdf in Figure \ref{fig:modelB4_triangle}. The halo mass and concentration are calculated from the the best-fitting $r_{\rm s}$ and $v_{200}$.}
\label{tab:modelB4_bestfit}
\end{table*}
%============================================

%--------------------------------------------
\subsection{JAM with an NFW dark matter halo} \label{sec:results_JAM_NFW}
%--------------------------------------------

%--------------------------------------------
The modelling attempts in the previous sections suggest that J1331's mass distribution could be more roundish in the inner regions and more massive at larger radii than expected from the distribution of stars alone. A DM halo in addition to the stellar component could explain these findings. 
%--------------------------------------------

%============================================
\begin{figure}
\centering
\includegraphics[width=\linewidth]{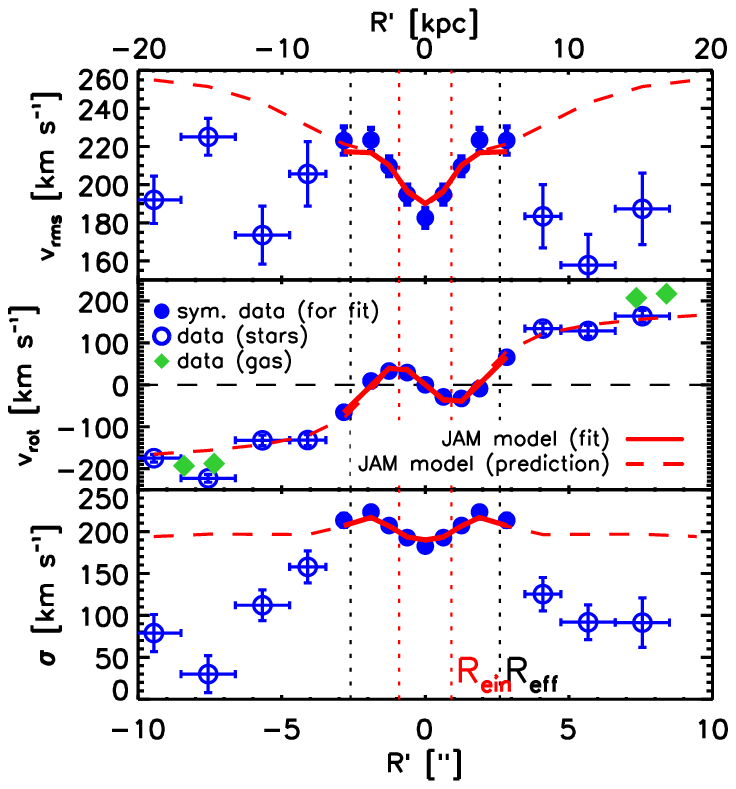}
\caption{Generating the rotation curve $v_\text{rot}(R')$ (red solid line in middle panel) from the best-fitting JAM model with NFW halo by fitting the rotation parameter $\kappa'$ to the symmetrized $v_\text{rot}$ data within $R'\sim3.5''$ (solid blue dots in central panel). The first panel shows the $v_\text{rms}$ data and model analogous to Figure \ref{fig:modelB4_vrms} for the mean parameters in Table \ref{tab:modelB4_bestfit}. The second panel shows the corresponding $v_\text{rot}$ for the best-fitting rotation parameter $\kappa' = 0.76$ in Equation \eqref{eq:kappa_profile} (red solid lines). The third panel shows simply the velocity dispersion $\sigma = \sqrt{v_\text{rms}^2 - v_\text{rot}^2}$. At larger radii we compare the unsymmetrized data (open blue dots), the gas kinematics from \citet{SWELLSV} (green diamonds) and an extrapolation of the JAM model, using the light distribution MGE from Figure \ref{fig:MGEouterRegions} for the outer regions (red dashed lines). The central regions are very well reproduced and we can also nicely predict the rotation curve at larger radii. Only at larger radii the $v_\text{rms}$ and $v_\text{rot}$ overestimate the measurements, probably due to a too massive NFW halo.}
\label{fig:modelB4_vrot}
\end{figure}
%============================================

%--------------------------------------------
\subsubsection{Modelling and priors}
%--------------------------------------------

%--------------------------------------------
We proceed by modelling the mass distribution with (i) a stellar component, which we get from the light MGE in Table \ref{tab:MGEF814W} (deprojected to the intrinsic $\nu(R,z)$ using the fixed inclination angle $\iota=70^\circ$ in Table \ref{tab:galaxyparameters}) times a constant stellar mass-to-light ratio $\Upsilon_\text{I,*}$, and (ii) a spherical NFW DM component (see Section \ref{sec:model_JAM_NFW}) with halo scale length $r_{\rm s}$ and circular velocity $v_{200}$ at the virial radius as free parameters. In the JAM modelling we use a 10-Gaussian MGE fit to the classical NFW profile from Equation \eqref{eq:NFWprofile}.

The full set of fit parameters is $(\Upsilon_\text{I,*},r_{\rm s},v_{200},\beta_z)$, where $\beta_z$ is again the constant velocity anisotropy parameter. We will investigate this parameter space with a Monte Carlo Markov Chain (MCMC; using code\footnote{The Python code package for \emph{emcee}, a Monte Carlo Markov Chain implementation by \citet{emcee}, is available online at \url{http://dan.iel.fm/emcee/current/}. The version from October 2013 was used in this work.} by\citealt{emcee}).

As we are particularly interested in the disentanglement of DM and stellar matter in the inner regions (see Section \ref{sec:intro}), we restrict the $v_\text{rms}$ fit to a region $R'\lesssim3.5''$, approximately within the effective half-light radius $R_\text{eff}=2.6''$. The constraints on a DM halo just from this data might be weak. To guide the fit towards a realistic NFW halo shape, we therefore impose several priors on the NFW halo.

\citet{Dutton10} give a relation for halo vs. stellar mass for late-type galaxies. Using the stellar mass estimate for J1331 from \citet{SWELLSI}, $m_* = (1.06 \pm 0.25) \times 10^{11} \text{M}_\odot$ (with generous error; see also Table \ref{tab:previousresults}) for the Chabrier IMF estimate, we find ${v_{200}} = (202_{-33}^{+44})_{-13}^{+12}$. The first error is due to the $2\sigma$ scatter in the relation by \citet{Dutton10}. The second error is the propagated error due to the uncertainty in the stellar mass. We use this as a rough estimate for the halo of J1331 and as Gaussian prior on $v_{200}$, 
\begin{equation}
p(v_{200}) = \mathscr{N}(200~\text{km s}^{-1},40~\text{km s}^{-1}). \label{eq:prior_v200}
\end{equation}

We also use the concentration vs. halo mass relation by \citet{Maccio08} in Equation \eqref{eq:Maccio08} as a prior on the concentration, i.e.
\begin{equation}
p(\log c_{200} \mid v_{200}) = \mathscr{N}\left(\langle \log c_{200} \rangle (M_{200}) , 0.105 \right). \label{eq:prior_c200}
\end{equation}

For the velocity anisotropy parameter $\beta_z$ we will again employ a uniform prior 
\begin{equation}
p(\beta_z) = \mathscr{U}(-0.5,+0.5)
\end{equation}
to exclude very unrealistic anisotropies. The full prior used is then
\begin{align}
\begin{split}
&p(\Upsilon_\text{I,*},r_{\rm s},v_{200},\beta_z) \\
&= \frac{1}{\ln\left( 10 r_{\rm s}\right)} p(\log c_{200} \mid v_{200}) \times p(v_{200}) \times p(\beta_z), 
\end{split}
\end{align}
where the factor $1/\ln\left( 10 r_{\rm s}\right)$ is the Jacobian of the transformation from the halo parameters $(v_{200},\log c_{200})$ to the fit parameters $(v_{200},r_{\rm s})$.

We also include the Einstein mass $M(<R_\text{ein})$ with a 10\% error as an additional constraint in the fit.
%--------------------------------------------

%--------------------------------------------
\subsubsection{Best-fitting JAM model} \label{sec:results_JAM_NFW_results}
%--------------------------------------------

%--------------------------------------------
Figure \ref{fig:modelB4_triangle} shows the posterior probability distribution (pdf) of the fit sampled with an MCMC. Overplotted are also the priors used to constrain the NFW halo. The mean parameters are summarized in Table \ref{tab:modelB4_bestfit}. We find that the best-fitting NFW halo strives to be more massive and with a higher concentration (due to a smaller $r_{\rm s}$) than  proposed by the priors. The model also prefers  very negative velocity anisotropies. Both, the high halo concentration and low $\beta_z$, are needed to reproduce the central dip of the $v_\text{rms}$ curve. Figure \ref{fig:modelB4_models} illustrates the range of best-fitting models according to the extent of the pdf. The models fit the $v_\text{rms}$ data in the inner regions quite well (Figure \ref{fig:modelB4_vrms}) and are also consistent with the Einstein mass (see Figure \ref{fig:modelB4_enclMass}). The extrapolation of the model to larger radii however overestimates the data, does not exhibit a drop around $\sim 6''$ at all and seems to be therefore overall too massive.
%--------------------------------------------

%--------------------------------------------
\subsubsection{Rotation curve}
%--------------------------------------------

%--------------------------------------------
We generate a rotation curve from the best-fitting mean model in Table \ref{tab:modelB4_bestfit}, whose $v_\text{rms}$ curve is shown in the first panel of Figure \ref{fig:modelB4_vrot}. Following the procedure in Section \ref{sec:model_JAM_rotation}, we find the rotation curve by fitting the rotation parameter $\kappa'$ to the symmetrized $v_\text{rot}$ data within $R' = 3.5''$. The best fit with $\kappa' = 0.76$ is given in the second panel of Figure \ref{fig:modelB4_vrot}. The third panel shows the dispersion that follows from $\sigma = \sqrt{v_\text{rms}^2 - v_\text{rot}^2}$. Our assumptions for $\kappa(R)$ nicely reproduce a $v_\text{rot}$ model with counter-rotating core. Although we fitted $v_\text{rot}$ only to the inner regions, the extrapolation to large radii matches the data also very well.

While the dispersion $\sigma$ in the center fits by construction quite good, the extrapolated dispersion is much larger than the data. We would expect the disk to be rotationally supported and therefore have a low velocity dispersion. Especially dispersions as high as $\sim 200~\text{km s}^{-1}$ are more likely to be observed in the pressure supported bulges of galaxies. There might be something unexpected with the $\sigma$ measurements around $\sim 5''$, but at large radii the best-fitting model NFW halo is simply too massive.

%--------------------------------------------

\subsubsection{Further tests and discussion}

We also fitted a model with a cored logarithmic DM halo. The cored halo models are in general slightly less massive than the NFW halo and therefore fit the outer regions of J1331 better. However, the density profile of the cored halo as well as the $I$-band light distribution within the plane drop as $\rho(r) \propto r^{-2}$. There is therefore a strong degeneracy between the stellar mass and the DM. Overall, we were not able to obtain tight constraints on the cored logarithmic halo.

As we only fitted the halo models to the inner regions, it is not surprising that they do not fit the kinematics at larger radii. A fit to all nine available $v_\text{rms}$ data points would force the DM halo to be less massive. But with decreasing DM contribution in the inner regions the model approaches the mass-follows-light model in the center that we already ruled out in Section \ref{sec:results_JAM_SB_MfL}. Allowing for free $\Upsilon_\text{I,*}(R')$, $\beta_z(R')$, and more flexible $\rho_\text{halo}(r)$ profiles, which would be the reasonable next step, would over fit the data. Without further priors on the stellar $\Upsilon_\text{I,*}(R')$ or more data points, we can make only the following statement: A model with constant $\Upsilon_\text{I,*} = 4.2\pm 0.2$, moderately tangential $\beta_z = -0.4\pm 0.1$ and a spherical and massive DM component with high concentration ($c_\text{200}=8\pm 2$) can explain the central kinematics of J1331 very well. We have however ruled out that in this case the DM halo will follow an NFW halo profile at both small and large radii.

We will also discuss the kinematics at larger radii in more depth in the discussion Section \ref{sec:discussion_kinematics}.

%--------------------------------------------
\section{Discussion and Conclusion} \label{sec:Discussion}
%--------------------------------------------

%--------------------------------------------
We have presented different dynamical models for the central region of J1331. Some of them capture the observed kinematics, but none of them work at both small and large radii. In the following we try to resolve some of the ambiguities by discussing possible reasons, by comparing our results to previous work and by hazarding some guesses on the true nature of J1331's matter distribution, which should be easily testable by future observations.
%--------------------------------------------

%============================================
\begin{table*}
\centering
\caption{Total $I$-band luminosity, stellar mass, and mass-to-light ratio, calculated from the $I$-band AB magnitudes and stellar masses found for J1331's bulge and disk by \citet{SWELLSI} (their table 2) for comparison with this work. The transformation from AB magnitudes to the Johnson-Cousins $I$-band used the relation $I[\text{mag}] = I[\text{ABmag}] - 0.309$ from \citet{FG1994} (their table 2). For the conversion from apparent magnitude to total luminosity the redshift $z=0.113$ \citet{SWELLSIII} was turned into a luminosity distance using the WMAP5 cosmology by \citet{WMAP5cosm}. }
\begin{tabular}{cccccc}
\hline\hline
& & \multicolumn{2}{c}{Chabrier IMF} & \multicolumn{2}{c}{Salpeter IMF}\\
      &  $L$ [$10^{10}L_{\odot}$]                & $M_*$ [$10^{10}\text{M}_\odot$]               & $\Upsilon_\text{I,*}^\text{chab}$ & $M_*$ [$10^{10}\text{M}_\odot$] & $\Upsilon_\text{I,*}^\text{sal}$ \\\hline
bulge &   $3.10 \pm 0.15 $  & $7.8 \pm 1.8$ & $2.5 \pm 0.6$ & $14.5 \pm 3.7 $ & $4.7 \pm 1.2$ \\
disk  &   $2.35 \pm 0.11 $  & $2.9 \pm 0.7$ & $1.2 \pm 0.3$ & $5.2 \pm 1.1$ & $2.2 \pm 0.5$ \\
total &   $5.45 \pm 0.19$ & $10.6 \pm 1.9$& & $19.7 \pm 3.9$&\\\hline
\end{tabular}
\label{tab:previousresults}
\end{table*}
%============================================

%--------------------------------------------
\subsection{On J1331's possible merger history} \label{sec:discussion_merger}
%--------------------------------------------

%--------------------------------------------
J1331 has a large counter-rotating stellar core within $\sim 2''$. This suggests a process in J1331's past in which two components with angular momenta oriented in opposite directions were involved.

Accretion of gas on retrograde orbits and subsequent star formation could lead to a younger and counter-rotating stellar population. However, to form enough stars such that the net rotation of the large and massive core is retrograde, a very large amount of gas would have had to be accreted by J1331---which is not very likely. 

Galaxy mergers are another possible scenario. Major mergers can form kinematically decoupled cores (KDCs) (e.g., \citealt{2011MNRAS.414.2923K,2015ApJ...802L...3T}), if they include large amounts of gas \citep{2010ApJ...723..818H}. During a minor merger, the dense nucleus of a satellite galaxy on a retrograde orbit could survive the dissipationless accretion and spiral to the galaxy's core due to tidal friction (e.g., \citealt{1984ApJ...287..577K,1988ApJ...327L..55F}). 

Usually ellipticals and the bulges of massive spirals appear reddest in their center and get increasingly bluer with larger radii. Mergers can reverse this behaviour by inducing the creation of young stellar populations in the remnant's center. Major mergers can trigger star formation bursts (e.g., \citealt{2008gady.book.....B}, \S 8.5.5). After a minor merger the satellite's stellar population now residing in the remnant's core is in general younger than the massive progenitor's bulge \citep{1996AJ....112..839C,2010MNRAS.404.1775T}. The different stellar populations in a merger remnant can be associated with different $\Upsilon_*$ and in rare cases they even show up as a reverse colour gradient within the bulge in photometry \citep{1990ApJ...361..381B, 1997ApJ...481..710C}.

Even though investigation of the photometry of J1331 did not reveal a distinct blue core in J1331, we cannot fully exclude the possibility that J1331 has such a $\Upsilon_*$ gradient (see discussion in Sections \ref{sec:results_JAM_SB_MfL} and \ref{sec:results_JAM_SB_gradient}). Spatially resolved stellar population analysis based on integral-field spectroscopy of J1331's bulge could provide further information on the true $\Upsilon_*(R')$.

Another way how galaxy encounters can modify the structure of galaxies is the excitation of warps \citep{1991wdir.book.....C,2013pss5.book..923S}. Warps lead to a twist in the projected kinematic major axis with radius, which is then also misaligned with the photometric major axis (\citealt{1998gaas.book.....B}, \S 8.2.4).

From kinematics along the photometric major axis only it can however not be determined if such a twist or misalignment is present in J1331, but it should be immediately visible in a 2D kinematic map.
%--------------------------------------------

%--------------------------------------------
\subsection{On J1331's central stellar mass-to-light ratio} \label{sec:MLdiscussion}
%--------------------------------------------

%--------------------------------------------
Some of the ambiguities in recovering J1331's matter distribution could be resolved by learning more about stellar populations with different IMFs in J1331. In particular, a sophisticated guess for the stellar mass-to-light ratio in the bulge could be compared to our very reliable measurement of the total mass-to-light ratio inside the Einstein radius $\Upsilon_\text{I,tot}^\text{ein} = 5.56 \Upsilon_{I,\odot}$. This would then either support or contradict the presence of a significant amount of DM in the bulge.

Traditional choices for the IMF are the bottom-heavy IMF by \citet{Salpeter1955}, $\xi(m) \propto m^{-x}$ with $x=2.35$, where $\xi(m) \diff m$ is the number of stars with mass $m$ in $[m,m+\diff m]$, and the IMFs by \citet{2002Sci...295...82K} and \citet{Chabrier2003}, which are in agreement with each other and predict less low-mass stars.

In the following we discuss why we think---based on our results and previous analyses---that J1331's bulge has an IMF slightly less bottom-heavy than the Salpeter-like IMF.

\emph{(i) Indications for a slightly less bottom-heavy Salpeter-like IMF in J1331's bulge:}

\citet{Ferreras} found a relation between the central stellar velocity dispersion $\sigma_0$ in early-type galaxies and the IMF slope $x$, where a higher $\sigma_0$ suggests a more bottom-heavy IMF. For a unimodal (Salpeter-like) IMF and $\sigma_0 \simeq 200~\text{km s}^{-1}$ in J1331 (see Figure \ref{fig:kinematics}), this relation predicts $x \approx 2.33$, which is close to the standard Salpeter slope, also supported by \citet{2014MNRAS.438.1483S}. When assuming a bi-modal (Kroupa-equivalent-like) IMF, \citet{Ferreras} predict $x \approx 2.85$ for J1331's central velocity dispersion. This is more bottom-heavy than the standard \citet{2002Sci...295...82K} IMF. Overall, the central velocity dispersion suggests a rather bottom-heavy IMF in J1331's bulge and therefore large stellar mass-to-light ratio. 

\citet{SWELLSI} estimated J1331's stellar bulge mass given a Salpeter IMF and measured the $I$-band AB magnitude of the bulge. Transformed to a stellar $I$-band mass-to-light ratio, their results would correspond to $\Upsilon_\text{I,*}^\text{sal} = 4.7 \pm 1.2$ (see Table \ref{tab:previousresults}). This is not too far from $\Upsilon_\text{I,*} = 4.2 \pm 0.2$ (see Table \ref{tab:modelB4_bestfit}), which we found when including an NFW halo in the JAM modelling.

\emph{(ii) Indications for and arguments against a Chabrier-like IMF in J1331's bulge:}

When \citet{SWELLSI} assumed a Chabrier IMF, their result translates to $\Upsilon_\text{I,*}^\text{chab} = 2.5 \pm 0.6$ (see Table \ref{tab:previousresults}). In Section \ref{sec:results_JAM_SB_gradient}, we created a dynamical model from only the surface brightness distribution and an increasing mass-to-light ratio profile without additional DM halo and without velocity anisotropy. We found that such a model would be perfectly consistent with the Einstein mass, predict a total $\Upsilon_\text{I,tot}(R'\sim0) = 2.53$---being consistent with the Chabrier IMF estimate by \citet{SWELLSI}---and rise quickly to $\Upsilon_\text{I,tot}(R'\gtrsim R_\text{ein}) \gtrsim 6$. 

This rise in $\Upsilon_\text{I,*}(R')$ could be either due to a cuspy DM halo (see Section \ref{sec:results_JAM_NFW}) or a very strong gradient in $\Upsilon_\text{I,*}$ (see Section \ref{sec:discussion_merger}). However, we rule out that a DM cusp is the sole reason because a cuspy and therefore overall massive DM halo does not match the kinematics in J1331's outer regions. We also rule out that a very strong $\Upsilon_\text{I,*}$ gradient is the only reason, because---as mentioned in Section \ref{sec:discussion_merger}---photometry did not reveal the clear existence of a blue population in the very center of J1331's bulge. Also, our modelling attempts allowing for velocity anisotropy (Sections \ref{sec:results_JAM_SB_MfL} and \ref{sec:results_JAM_NFW}) suggest that we do need some moderate $\beta_z<0$ to explain the central $v_\text{rms}$ dip. And lastly---as laid out in the previous paragraph---J1331's central velocity dispersion suggests a more bottom-heavy IMF. Overall it is therefore not very likely that the central regions of J1331 have such a low $\Upsilon_\text{I,*}^\text{chab} \sim 2.5$.

\emph{(iii) Arguments against an IMF more bottom-heavy than the Salpeter IMF in J1331's bulge:}

We also compare our results from Section \ref{sec:results_JAM_NFW} with the study by \citet{SWELLSV}. They found that the bulge of J1331 has an IMF \emph{more} bottom-heavy than the Salpeter IMF. Our fitting attempt---using more data within $\sim 3.5''$ than \citet{SWELLSV}---in Section \ref{sec:results_JAM_NFW} gave $\Upsilon_\text{I,*} = 4.2 \pm 0.2$ as best fit (see Table \ref{tab:modelB4_bestfit}), which indicates a \emph{less} bottom-heavy IMF than the Salpeter IMF. In Section \ref{sec:discussion_kinematics}, we will argue why we do not think that the \cite{SWELLSV} model is a good model for the central regions of J1331.
%--------------------------------------------

%--------------------------------------------
\subsection{On J1331's kinematics} \label{sec:discussion_kinematics}
%--------------------------------------------

%--------------------------------------------
Overall, the kinematics of the merger remnant J1331 are peculiar. In particular, there are two features in the $v_\text{rms}$ curve that are hard to explain with standard modelling techniques. The first feature, the deep central $v_\text{rms}$ dip, was studied in detail in this work. The second feature, the drop and rise in $v_\text{rms}$ around $R'\sim6''$, is even harder to explain; below we will speculate about possible reasons that could cause such a signature.

\emph{(i) The central $v_\text{rms}$ dip within $R'\lesssim1''$:}

\begin{figure}
\centering
\includegraphics[width=0.9\linewidth]{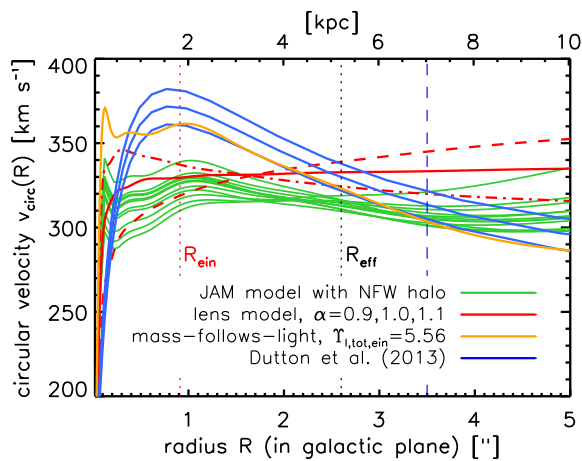}
\caption{Comparison of the circular velocity curve of J1331's inner regions for different models: (i) JAM model with NFW DM halo from Section \ref{sec:results_JAM_NFW_results} and Figure \ref{fig:modelB4_models} (green). (ii) Lens model from Section \ref{sec:results_lensing_bestfit} and Figure \ref{fig:JAM_modelL} with $\alpha = 1.1$ (red dashed line), $\alpha = 1$ (red solid line) and $\alpha=0.9$ (red dash-dotted line). (iii) Mass-follows-light model, which uses the F814W surface brightness in Table \ref{tab:MGEF814W} and the mass-to-light ratio in the Einstein radius, $\Upsilon^\text{ein}_\text{I,tot} = 5.56$, to generate a mass distribution, as in Figure \ref{fig:lenscomparelight} (orange line).  (iv) Model from gas kinematics and Einstein mass found by \citet{SWELLSV} (their Figure 2, best model with 68\% confidence region) (blue lines).}
\label{fig:vcirc_comparison}
\end{figure}

First, we will compare our modelling results with the modelling results by \citet{SWELLSV} within $R_\text{eff}$. \citet{SWELLSV} fitted a stellar mass model and NFW halo to (i) the Einstein mass and (ii) gas kinematics at larger radii $\gtrsim8''$. Figure \ref{fig:vcirc_comparison} compares the circular velocity curve found by \citet{SWELLSV} with a mass-follows-light model scaled to fit our Einstein mass (by multiplying the light distribution in Table \ref{tab:MGEF814W} by $\Upsilon_\text{I,tot}^\text{ein} = 5.56$, analogous to Figure \ref{fig:lenscomparelight}). Within $0''.5 < R < 5''$ they agree with each other. 

The models in this work used more than just the Einstein mass to constrain the matter distribution at small radii: The lens mass model constrained also the shape of the mass distribution within the lensing image configuration at $R_\text{ein} \sim 1''$. The dynamical models used stellar kinematics inside $R' \simeq 3.5''$. We compare the lens mass model's $v_\text{circ}$ (for $\alpha=1.0\pm 0.1$) with the NFW JAM model (Table \ref{tab:modelB4_bestfit}) in Figure \ref{fig:vcirc_comparison} as well. Within and around $R_\text{ein}$ they are consistent with each other, even though they were independently derived. They do not agree with the mass-follows-light-like result by \citet{SWELLSV} and in Section \ref{sec:results_JAM_SB_MfL} we showed that ``mass-follows-light'' is not a good model for J1331.

%============================================
%============================================

Because we used more data constraints in the center than \citet{SWELLSV}, we think that our model for J1331's bulge is more reliable. In conclusion, we suspect that the most likely model for J1331's central bulge is a moderate DM contribution in the center (Section \ref{sec:results_JAM_NFW_results}) with some tangential anisotropy (Sections \ref{sec:results_JAM_SB_MfL} and \ref{sec:results_JAM_NFW}, which would be also consistent with the counter-rotation in the bulge) and an overall $\Upsilon_\text{*,I}$ in the bulge which is slightly lower than that of a Salpeter-like IMF (Section \ref{sec:MLdiscussion}). Different stellar populations inside the bulge due to the merger (Section \ref{sec:discussion_merger}) with different $\Upsilon_{*}$ could add to an overall rising $\Upsilon_\text{tot}(R')$ profile (Sections \ref{sec:results_lensing_compare} and \ref{sec:results_JAM_SB_gradient}).

%--------------------------------------------

\emph{(ii) The drop and rise of the $v_\text{rms}$ curve around $R'\sim 6''$:}

The drop in $v_\text{rms}$ around $R'\sim4''$ (Figure \ref{fig:kinematics}) is most likely due to the transition from red bulge to blue disk (compare bulge size in Figures \ref{fig:F450W} and \ref{fig:F814W}). A mix of different stellar populations in the disk could make the modelling difficult: The smooth light distribution was derived from the F814W filter and is therefore dominated by older stars, while the measured light-weighted stellar kinematics in the disk are dominated by luminous young stars with their lower velocity dispersion. As the merger could have caused a warp in J1331's disk, it is possible that bulge and disk have different inclination angles and a kinematic twist. The latter could have lead to a misalignment of the long-slit and the galaxy's kinematic major axis at larger radii, therefore to measurements of lower $v_\text{rot}$ and consequently to a stronger underestimation of the true $v_\text{circ}$, which would add to the drop in $v_\text{rms}$.

Around $R'\sim7''$ the rotation curve and dispersion show some wiggles which lead to a spontaneous rise of the $v_\text{rms}$. As the spiral arms with their non-circular motions and patchy star-forming regions cross the major axis around this radius, we suspect they cause this disturbance of the axisymmetric kinematics.

\emph{(iii) The profile of the DM halo at $R'\gtrsim 5''$:}

The DM halo should start to dominate the kinematics at larger radii. \cite{SWELLSV}, who fitted an NFW halo to the gas kinematics in the outer regions, found lower halo masses ($v_\text{circ,halo}(5'') \sim 120~\text{km s}^{-1}$ according to their Figure 2) than we did ($v_\text{circ,halo}(5'') \sim 200~\text{km s}^{-1}$, Figure \ref{fig:vcirc_comparison}). As we did not fit the outer regions and only used a prior for $v_\text{200}$, their result in this regime is more reliable. Given our findings that an NFW halo does not fit the kinematics at both small and large radii, we suspect that the true halo has a profile that deviates strongly from an NFW halo, possibly as a result of the merger.

All of these speculations should be easily testable with 2D kinematics.

%--------------------------------------------
\subsection{Future work}
%--------------------------------------------

%--------------------------------------------
J1331's merging history and peculiar kinematics make it a valuable and exiting target to study merger remnants and a challenge for dynamical modelling techniques. We found however that the existing data alone---photometry and major axis kinematics---is not sufficient to resolve all the ambiguities we encountered in our modelling.

The main future work would be therefore getting high-resolution integral-field spectroscopy for J1331. High spatial resolution would be required to clearly identify J1331's presumably complex kinematic structure. High spectral resolution would be important to be able to reliably measure the low velocity dispersion in the outer regions of J1331.

Specifically 2D kinematics should help to answer the following questions: Is the drop in $v_\text{rms}$ around $R' \sim 3-6~\text{kpc}$ real? Did the long slit spectrograph maybe miss the major axis of the disk? And most importantly: Are the kinematics asymmetric? Is it possible that there even exists a kinematic twist due to the merger in J1331's past? In the latter two cases we would need to apply non-axisymmetric Jeans models to J1331 as the assumption of axisymmetry of this work would not be valid anymore.

Furthermore, learning more about different stellar populations in J1331 would lead to valuable constraints for the modelling. While a quick investigation of the photometric colour profile did not reveal obvious colour gradients in J1331's bulge, there could be still stellar $\Upsilon_{I,*}$ variations due to age or metallicity differences. Existing major axis spectroscopy and/or future IFU data could be employed (i) to investigate absorption line indices to confirm (or contradict) the existence of $\Upsilon_{I,*}$ gradients and (ii) to perform stellar population analyses to constrain $\Upsilon_{I,*}$ reliably and independently of kinematics.

Future modelling approaches should fit dynamics (stellar and gas kinematics) simultaneously with the gravitational lensing (image positions, shape and even flux ratios) in a similar fashion to \citet{SWELLSIV}. To also model the extent, shape and flux of the lensing images, the method by \citet{2004ApJ...611..739T,2003ApJ...590..673W} could be employed, which models the surface brightness distribution of the images and source on a pixelated grid. For this to work, a good model for the galactic extinction would be needed---but 2D spectroscopy could also help with this.

All of the above would lead to a much better understanding of J1331's structure and mass distribution and therefore answer questions on how mergers might modify spiral galaxies.
%--------------------------------------------

%--------------------------------------------
\subsection{Summary}
%--------------------------------------------

%--------------------------------------------
We constrained the matter distribution of the massive spiral galaxy J1331, which has a large counter-rotating core, probably due to a merger in its past, and acts as a strong gravitational lens for a background source. We used two independent methods to model J1331: gravitational lensing and dynamical Jeans modelling. We focused on the bulge region of J1331 to complement previous studies of J1331 by \citet{SWELLSIII} and \citet{SWELLSV}. The mass constraints from lensing and dynamics agree very well with each other within $R_\text{eff}$.

In our lensing approach we fitted a scale-free galaxy model to the lensing image position. This constrained the Einstein radius to within 2\% [$R_\text{ein}=(0.91\pm0.02)'' \hat{=}(1.83\pm0.04)~\text{kpc}$], and the Einstein mass to within 4\% ($M_\text{ein} = (7.8\pm0.3) \times 10^{10} \text{M}_\odot$), consistent with results by \citet{SWELLSIII}.

A MGE fit to J1331's surface brightness in the F814W filter in HST imaging by \citet{SWELLSI} helped us determining the effective radius, $R_\text{eff} \simeq 2.6'' \hat{=} 5.2~\text{kpc}$, and total $I$-band luminosity of the galaxy, $L_\text{I,tot} \simeq 5.6 \times 10^{10} L_\odot$.

Axisymmetric JAM modelling allowed a comparison between model predictions for the second velocity moment given a tracer and mass distribution and the observed stellar kinematics from major axis long slit spectroscopy by \citet{SWELLSV}. The independent JAM model of the lens mass model was consistent with observed kinematics within $R_\text{eff}$. We also fitted mass models with and without NFW halo to the stellar kinematics within $R'=3.5~\text{kpc}$. From this we deduced that a mass-follows-light model (even with velocity anisotropy) is not a good model for J1331's inner regions. This ruled out the previous findings of \citet{SWELLSV} for J1331's bulge. We discussed that, to describe the observed stellar kinematics, we most likely require a moderate contribution of a DM halo already in the bulge, moderate tangential velocity anisotropy, $\beta_z \simeq -0.4\pm0.1$, and possibly even a varying stellar mass-to-light ratio, which could be the result of the previous merger event. We argue that we expect the total stellar mass-to-light ratio within the bulge to be $\Upsilon_\text{I,*}\simeq 4.2\pm0.2$, which is slightly less bottom-heavy than a Salpeter IMF ($\Upsilon_\text{I,*}^\text{sal}\sim 4.7$). We also showed that it is possible to construct a model which includes the counter-rotating core and fits the rotation curve at large radii.

While both our independent mass models are consistent with each other within $\sim R_\text{eff}$, they do not describe the data at large radii very well. We speculate how a merger could have modified the kinematic structure and/or mass distribution of J1331. To resolve the ambiguities in J1331's mass distribution two-dimensional kinematic maps of J1331 from integral-field unit spectroscopy are needed.
%--------------------------------------------

\section*{Acknowledgements}

The authors thank Tommaso Treu and the SWELLS team for their support, and the anonymous referee for their helpful comments. WHT thanks the Department of Physics and Astronomy at the University of Heidelberg for awarding this research with the Otto-Haxel-Prize for an outstanding master's thesis. GvdV acknowledges partial support from Sonderforschungsbereich SFB 881 "The Milky Way System" (subproject A7 and A8) funded by the German Research Foundation, and from the DAGAL network from the People Programme (Marie Curie Actions) of the European Union's Seventh Framework Programme FP7/2007-2013/ under REA grant agreement number PITN-GA-2011-289313.

%--------------------------------------------
\bibliography{literaturelist}{}
\bibliographystyle{mnras}
%--------------------------------------------

\label{lastpage}

\end{document}